\documentclass[%
 aip,
 jmp,%
 amsmath,amssymb,
preprint,%
nofootinbib
]{revtex4-1}

\usepackage{graphicx}
\usepackage{color}
\usepackage{bm}
\usepackage{amssymb}
\usepackage{tikz}
\usepackage{multirow}
\usepackage{epstopdf}
\usepackage{natbib}

\usepackage{algorithm}
\usepackage{algpseudocode}
\usepackage{amsmath,amsfonts,amsthm} 
\usepackage{verbatim} 

\usepackage{multirow}
\usepackage{epstopdf}
\usepackage{natbib}

\def\nips#1#2#3{{\it NIPS},~pages~ #1-#2,~(#3)}

\begin{document}


\title{Physical extrapolation of quantum observables by generalization with Gaussian Processes}


\author{R. A. Vargas-Hern\'{a}ndez and R. V. Krems}
\affiliation{\scriptsize 
Department of Chemistry, University of British Columbia, Vancouver, BC V6T 1Z1, Canada}


\date{\today}

\begin{abstract}
For applications in chemistry and physics, machine learning is generally used to solve one of three problems: interpolation, classification or clustering. 
These problems use information about physical systems in a certain range 
of parameters or variables in order to make predictions at unknown values of these variables within the same range. 
The present work illustrates the application of machine learning to prediction of physical properties outside the range of the training parameters.   
We define  `physical extrapolation' to refer to accurate predictions $y(\bm x^\ast)$ of a given physical property at a point $\bm x^\ast = \left [ x^\ast_1, ..., x^\ast_{\cal D} \right ]$ in the $\cal D$-dimensional space, if, at least, one of the variables $x^\ast_i \in \left [ x^\ast_1, ..., x^\ast_{\cal D} \right ]$ is {\it outside} of the range covering the training data. 
We show that Gaussian processes can be used to build machine learning models capable of physical extrapolation of quantum properties 
of complex systems across quantum phase transitions. 
The approach is based on training Gaussian process models of variable complexity 
by the evolution of the physical functions. We show that, as the complexity of the models increases, they become capable of predicting 
new transitions. We also show that, where the evolution of the physical functions is analytic and relatively simple (one example considered here is $a + b/x + c/x^3$), Gaussian process models with simple kernels already yield accurate generalization results, allowing for accurate predictions of quantum properties in a different quantum phase.  
For more complex problems, it is necessary to build models with complex kernels. The complexity of the kernels is increased using the Bayesian Information Criterion (BIC). We illustrate the importance of the BIC by comparing the results with random kernels of various complexity. 
We discuss strategies to minimize overfitting and illustrate a method to obtain meaningful extrapolation results without direct validation in the extrapolated region.

\end{abstract}


\maketitle

\section{Introduction}
\label{sec:intro}

As described throughout this book, machine learning has in recent years become a powerful tool for physics research. 
A large number of machine learning applications in physics can be classified as supervised learning, which aims to build a model ${\cal F}(\cdot)$
of the $\bm x \mapsto y$ relation,  given a finite number of $\bm x_i \mapsto y_i$ pairs. Here, $\bm x$ is a vector of (generally multiple) parameters determining the physical problem of interest and $y$ is a physics result of relevance. For example, $\bm x$ could be a vector of coordinates specifying the positions of atoms in a polyatomic molecule and $y$ the potential energy of the molecule calculated by means of a quantum chemistry method \cite{fitting-1,fitting-2,fitting-3,fitting-4,fitting-5,sergei-nn,JiePRL,JieJPhysB,VargasBO,GPvsNN,GPper}. In this case, ${\cal F}(\bm x)$ is a model of the potential energy surface constructed based on $n$ energy calculations $\bm y = \left ( y_1, ..., y_n \right )^\top$ at $n$ points $\bm x_i$ in the configuration space of the molecule. To give another example, $\bm x$ could represent the parameters entering the Hamiltonian of a complex quantum system (e.g., the tunnelling amplitude, the on-site interaction strength and/or the inter-site interaction strength of an extended Hubbard model) and $y$ some observable such as the energy of the system. Trained by a series of calculations of the observable at different values of the Hamiltonian parameters,  ${\cal F}(\bm x)$ models the dependence of the observable on the Hamiltonian parameters 
{\cite{WangPRB2016,CarasquillaNatPhys2017,  NieuwenburgNatPhys2017, BroeckerarXiv2017, WetzelScherzerarXiv2017,Wetzel2017,LiuarXiv2017,ChangPRX2017, BroeckerSciRep2017, SchindlerPRB2017, OhtsukiJPSJapan2016, ArsenaultPRB2014, ArsenaultarXiv2015, BeacharXiv2017, RafaelarXiv2017, YoshiokaarXiv2017, VenderleyarXiv2017, RBM_Troyer, SchmittarXiv2017, CaiPRB, HuangarXiv, DengPRB, NomuraarXiv2017, DengPRX, GaoNatComm, TorlaiarXiv2017, GPdeepNN_2015, DanielyNIPS2016, GPdeepNN_2017,MLcrystal1, MLcrystal2, MLcrystal3, MLcrystal4}}, which could be used to map out the phase diagram of the corresponding system. 

The ability of a machine learning model to predict previously unseen data is referred to as `generalization'. These previously unseen data must usually come from the same distribution as the training data, but may also come from a different distribution. Most of the applications of machine learning in physics aim to make predictions within the range of training data.  
In the present work, we discuss a method for building machine learning models suitable for physical extrapolation. We define  `physical extrapolation' to refer to accurate predictions $y(\bm x^\ast)$ of a given physical property at a point $\bm x^\ast = \left [ x^\ast_1, ..., x^\ast_{\cal D} \right ]$ in the $\cal D$-dimensional input space, if, at least, one of the variables $x^\ast_i \in \left [ x^\ast_1, ..., x^\ast_{\cal D} \right ]$ is {\it outside} of the range covering the training data. Thus, in the present work, the training data and test data distributions are necessarily separated in input space. We will refer to the predictions of machine learning models as generalization and the physical problems considered here as extrapolation. 

\begin{figure}[h!]
\centering
\includegraphics[width=0.7\columnwidth]{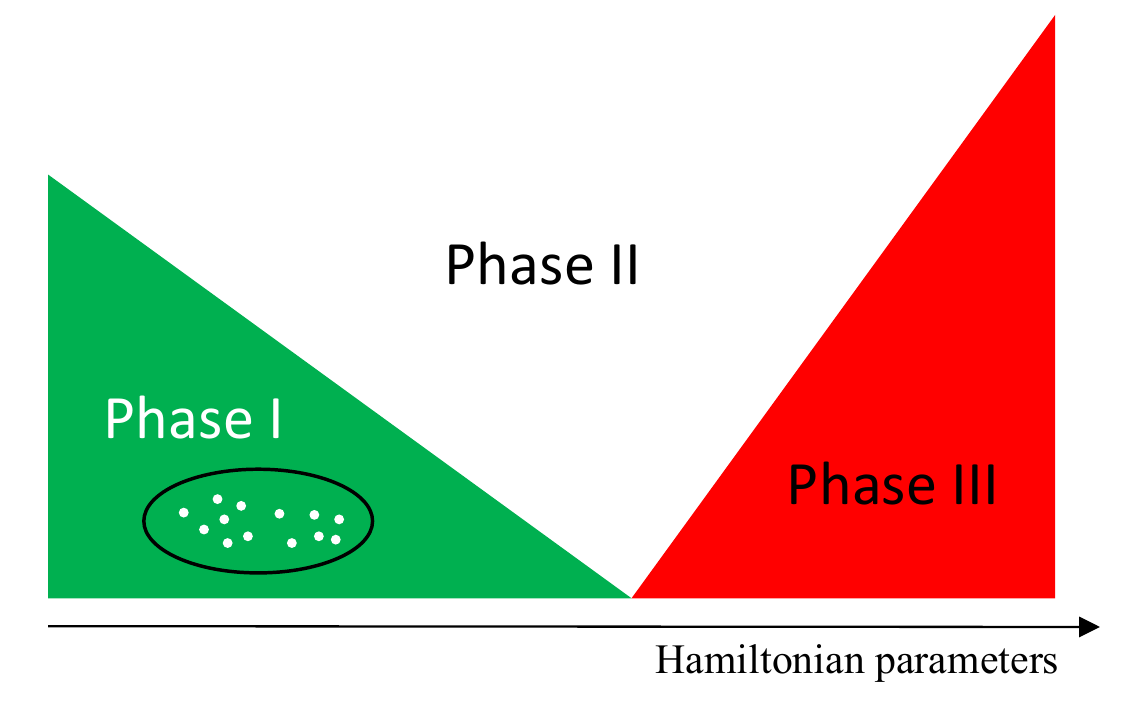}
\caption{Schematic diagram of a quantum system with three phases. The goal of the present work is to predict both of the phase transitions based on information about the properties of the system in the encircled region of phase I. }
\label{phases}
\end{figure}

Our particular goal is to extrapolate complex physical behaviour without {\it a priori} knowledge of the physical laws governing the evolution of the system. 
For this purpose,  we consider a rather challenging problem: prediction of physical properties of complex quantum systems with multiple phases
based on training data entirely in one phase. The main goal of this work is schematically illustrated in Figure \ref{phases}. We aim to construct the machine learning models that, when trained by the calculations or experimental measurements within one of the Hamiltonian phases (encircled region in Figure \ref{phases}), are capable of predicting the physical properties of the system in the other phases. Of particular interest is the prediction of the phase transitions, which are often challenging to find with rigorous quantum calculations. 

This problem is challenging because the wave functions of the quantum systems -- as well as the physical observables characterizing the phases -- undergo sharp changes at the phase transitions. Most of the machine learning models used for interpolation/fitting are, however, smooth functions of $\bm x$. So, how can one construct a machine learning model that would capture the sharp and/or discontinuous variation of the physical properties? The method discussed here is based on the idea put forward in our earlier work \cite{extrapolation-paper}.  

We assume that the properties of a physical system within a given phase contain information about multiple transitions and that,  when a system approaches a phase transition, the properties must change in a way that is affected by the presence of the transition as well as the properties in the other phase(s).  In addition, a physical system is often characterized by some properties that vary smoothly through the transition. The goal is then to build a machine learning model that could be trained by such properties within a given phase, make a prediction of these properties in a different  phase and predict the properties that change abruptly at the transition from the extrapolated properties. We will use Gaussian processes to build such models. 

\subsection{Organization of this chapter}
\label{sec:org}
The remainder of this chapter is organized as follows. The next section describes the quantum problems considered here. 
Understanding the physics of these problems is not essential for understanding the contents of this chapter. The main purpose of Section \ref{sec:quantum_sys} is to introduce the notation for the physical problems discussed here. These problems are used merely as examples.  Section \ref{sec:GP_interpolation} briefly discusses the application of Gaussian process regression for interpolation in multi-dimensional spaces, mainly to set the stage and define the notation for the subsequent discussion. Section \ref{sec:GP_extrapolation} describes the extension of Gaussian process models to the extrapolation problem. Section \ref{sec:extrap_quantum_prop} presents the results and Section \ref{sec:conclusion} concludes the present chapter. We will abbreviate `Gaussian process' as GP,  `Artificial Neural Network' as NN  and `machine learning' as ML throughout this chapter. 

\section{Quantum systems}
\label{sec:quantum_sys}

In this section, we describe the quantum systems considered in the present work. In general, we consider a system described by the Hamiltonian $\hat H = \hat H(\Gamma)$ that depends on a finite number of free parameters $\Gamma = \{ \alpha, \beta, \dots \}$. The observables depend on these Hamiltonian parameters as well as the intrinsic variables ${V} = \{ v_1, v_2, \dots\}$ such as the total linear momentum for few-body systems or thermodynamic variables for systems with a large number of particles. The set $\Gamma + {V}$ comprises the independent variables of the problems considered here.  The ML models $\cal F$ will be functions of $\Gamma + {V}$. 

More specifically, we will illustrate the extrapolation method using two completely different quantum models: the lattice polaron model and the mean-field Heisenberg model.

\subsection{Lattice polarons} 
\label{subsec:lattice_pol}

The lattice polaron model describes low-energy excitations of a quantum particle hopping on a lattice coupled to the bosonic field provided by lattice phonons. 
We consider a quantum particle (often referred to as the `bare' particle) in a one-dimensional lattice with $N \rightarrow \infty$ sites coupled to a phonon field: 
 \begin{eqnarray}
{\cal H} =   \sum_k  \epsilon_k c^\dagger_k c_k +  \sum_q \omega_q  b^\dagger_q b_q + V_{\rm{e-ph}},
\label{polaron}
\end{eqnarray}
where $c_k$ and $b_q$ are the annihilation operators for the bare particle with momentum $k$ and phonons with momentum $q$,  $\epsilon_k = 2 t \cos(k)$ is the energy of the bare particle and $\omega_q = \omega = {\rm const}$ is the phonon frequency. The particle-phonon coupling is chosen to represent a combination of two qualitatively different polaron models:
\begin{eqnarray}
V_{\rm{e-ph}} = \alpha H_1 + \beta H_2, 
\end{eqnarray}
where 
\begin{eqnarray}
H_1 = \sum_{k,q} \frac{2i}{\sqrt{N}}\left [ \sin(k + q) - \sin(k) \right]c^\dagger_{k+q}c_k \left ( b^\dagger_{-q}  + b_q\right ) \quad
\end{eqnarray}
describes the Su-Schrieffer-Heeger (SSH) \cite{ssh} particle-phonon coupling, and  
\begin{eqnarray}
H_2 = \sum_{k,q} \frac{2i}{\sqrt{N}} \sin(q)c^\dagger_{k+q}c_k \left ( b^\dagger_{-q}  + b_q\right )
\end{eqnarray}
is the breathing-mode model \cite{breathing-mode}. We will focus on two specific properties of the polaron in the ground state: the polaron momentum  and the polaron effective mass. 
The ground state band of the model (\ref{polaron}) represents polarons whose effective mass and ground-state momentum are known to exhibit two sharp transitions as the ratio $\alpha/\beta$ increases from zero to large values \cite{HerreraPRL}. At $\alpha = 0$, the model (\ref{polaron}) describes breathing-mode polarons, which have no sharp transitions \cite{no-transition}. At $\beta = 0$, the model (\ref{polaron}) describes  SSH polarons, whose effective mass and ground-state momentum exhibit one sharp transition in the polaron phase diagram \cite{ssh}. At these transitions, the ground state momentum and the effective mass of the polaron change abruptly.

\subsection{The Heisenberg model}
\label{subsec:Heis_model}

The second model we consider here is the Heisenberg model
\begin{eqnarray}
H = -\frac{J}{2}\sum_{\langle i,j \rangle}\vec{S}_i \cdot \vec{S}_j.
\end{eqnarray}
This model describes a lattice of interacting quantum spins $S_i$, which -- depending on the strength of the interaction $J$ -- can be either aligned in the same direction (ferromagnetic phase) or oriented randomly leading to zero net magnetization (paramagnetic phase).  The parameter $J$ is the amplitude of the interaction and the $\langle .. \rangle$ brackets indicate that the interaction is non-zero only between nearest neighbour spins.

Within a mean-field description, this many-body quantum system has free energy density \cite{Chaikin, sachdev}
\begin{eqnarray}
f(T,m) \approx \frac{1}{2}\left (1- \frac{T_c}{T} \right) m^2 + \frac{1}{12}\left (\frac{T_c}{T} \right)^3 m^4,
\label{mf_H}
\end{eqnarray}
where $m$ is the magnetization, $T$ is the temperature and $T_c$ is the critical temperature of the phase transition. At temperatures ${T} > T_c$, the model yields the paramagnetic phase, while ${T} < T_c$ corresponds to the ferromagnetic phase. The main property of interest here will be the order parameter. This property undergoes a sharp change at the critical temperature ovf the paramagnetic - ferromagnetic phase transitions.

\section{Gaussian process regression for interpolation}
\label{sec:GP_interpolation}

The purpose of GP regression is to make a prediction of some quantity $y$ at an arbitrary point $\bm x \in \left [ \bm x_{\rm min}, \bm x_{\rm max}  \right ]$ of a $\cal D$-dimensional space, given a finite number of values $\bm y = \left ( y_1, ..., y_n \right )^\top$, where $y_i$ is  the value of $y$ at $\bm x_i$. Here, $\bm x_i$ is a $\cal D$-dimensional vector specifying a particular position in the input space and it is assumed that the values $\bm x_i$ sample the entire range   $\left [ \bm x_{\rm min}, \bm x_{\rm max}  \right ]$. If the training data are noiseless (as often will be the case for data coming from the solutions of physical equations), it is assumed that $y$ is represented by a continuous function $f$ that passes through the points $y_i$, so the vector of given results is 
$\bm y = \left ( f(\bm x_1), ..., f(\bm x_n) \right )^\top$.  The goal is thus to infer the function $f(\bm x)$ that interpolates the points $y_i \equiv f(x_i)$. The values $y_i$ in the vector $\bm y$ represent the `training data'.  

\begin{figure}[h!]
	\includegraphics[width=0.99\columnwidth]{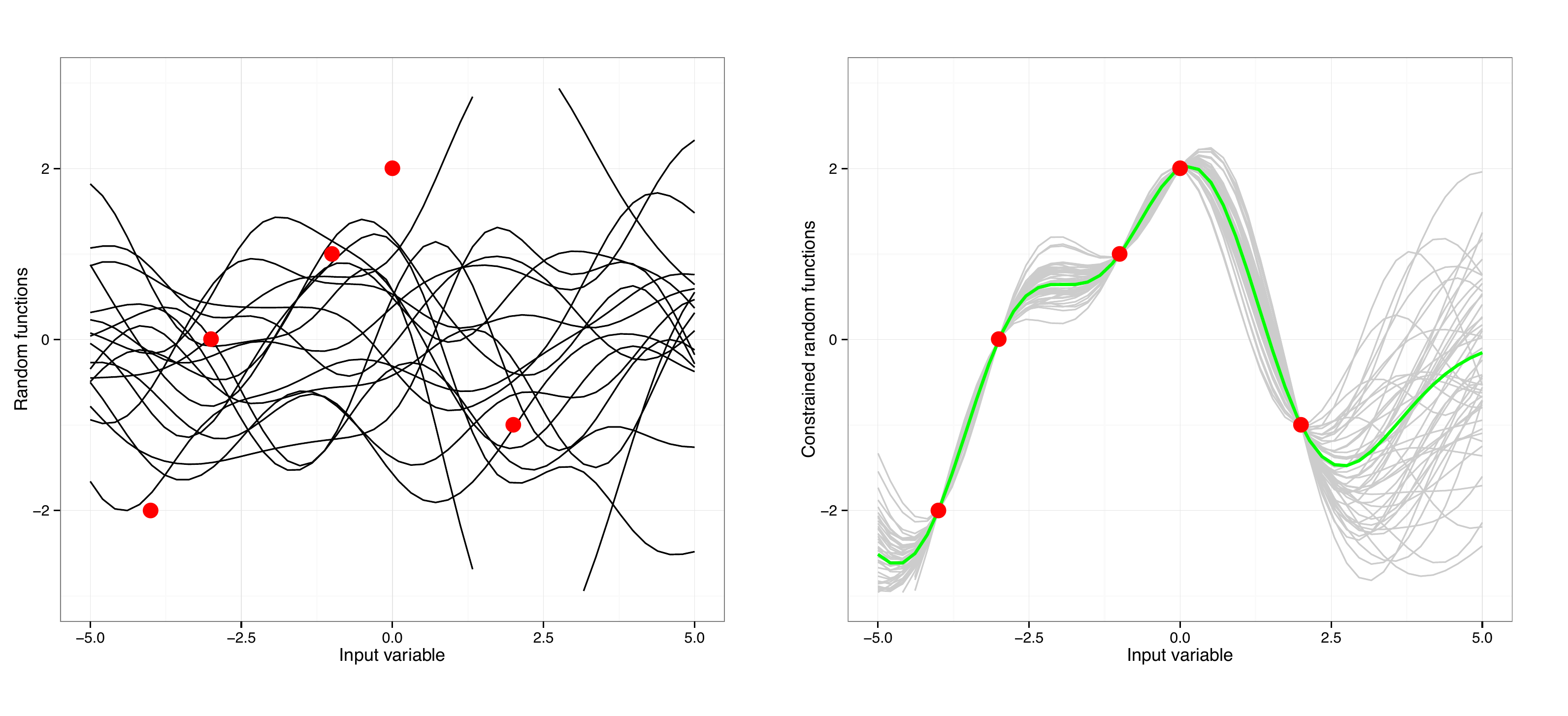}
	\caption{Left:  Gaussian process prior (grey curves). Right: Gaussian process (grey curves) conditioned by the training data (red dots). The green curve represents the mean of the GP posterior.  
	}
	\label{fig:GP_RK}
\end{figure}

GPs  infer a {distribution over functions} $p(f| {\bm y})$ given the training data, as illustrated in Figure \ref{fig:GP_RK}. 
The left panel of Figure \ref{fig:GP_RK} shows an example of the GP prior, i.e. the GP before the training. The right panel shows the GP conditioned by the training data (red dots). 
The GP is characterized by a mean function ${\bm \mu}({\bm x})$ and covariance $\Sigma({\bm x})$. The matrix elements of the covariance are defined as $\Sigma_{ij} = k({\bm x}_i,{\bm x}_j)$, where $k(\cdot,\cdot)$ is a positively defined kernel function.

It is possible to derive the closed-form equations for the conditional mean and variance of a GP \cite{gpbook}, yielding
\begin{eqnarray}
\mu(\bm x_*) &=& K({\bm x_*},\bm x)^\top \left [ K(\bm x, \bm x) + \sigma_n^2 I \right ] ^{-1}{\bm y} \label{eqn:gp_mu}
\label{meanGP}
\\
\sigma(\bm x_*) &=& K({\bm x_*},{\bm x_*}) - K({\bm x_*},\bm x)^\top\left [ K(\bm x,\bm x) + \sigma_n^2 I\right ]^{-1}K({\bm x_*},\bm x),
\label{varianceGP}
\end{eqnarray} 
where  $\bm x_\ast$ is a point in the input space where the prediction $\bm y_\ast$ is to be made;
 $K(\bm x, \bm x)$ is the $n \times n$ square matrix with the elements $K_{i,j} = k(\bm x_i,\bm x_j)$ representing the covariances between $y(\bm x_i)$ and $y(\bm x_j)$. The  elements $k(\bm x_i,\bm x_j)$ are represented by the kernel function. 
Eq. (\ref{meanGP}) can then be used to make the prediction of the quantity $y$ at point $\bm x^\ast$, while Eq. (\ref{varianceGP}) can be used to define the error of the prediction. 

In this work, the GP models are trained by the results of quantum mechanical calculations. For the case of the polaron models considered here,  

\begin{eqnarray}
\nonumber
\bm x_i \Rightarrow \{ {\rm polaron~momentum~}K,~ {\rm Hamiltonian~parameter}~\alpha, \\ ~ {\rm Hamiltonian~parameter}~\beta, {\rm phonon~frequency~}\omega\}. 
\nonumber
\end{eqnarray}

\noindent
For the case of the Heisenberg model considered here, {$$\bm x_i \Rightarrow \{ {\rm Temperature}~T,~ {\rm magnetization}~{m} \}$$}

As already mentioned, $\bm y \Rightarrow f(\bm x)$ is a vector of quantum mechanics results at the values of the parameters specified by $\bm x_i$.
For the case of the polaron models considered here, $\bm y \Rightarrow {\rm polaron~energy~}E$. For the case of the Heisenberg model considered here, 
$\bm y \Rightarrow {\rm free~energy~density}$.

To train a GP model, it is necessary to assume some analytic form for the kernel function $k(\cdot, \cdot)$. In the present work, we will use the following analytic forms for the kernel functions: 
\begin{eqnarray}
k_{\rm LIN}({\bm x}_i, {\bm x}_j)  &=&  {\bm x}_i^\top {\bm x}_j + {\ell}
\label{eqn:k_LIN}\\
k_{\rm RBF}({\bm x}_i, {\bm x}_j) &=& \exp \left(-\frac{1}{2}r^2({\bm x}_i, {\bm x}_j)\right)
\label{eqn:k_RBF}\\
k_{\rm MAT}({\bm x}_i, {\bm x}_j)  &=& \left( 1 + \sqrt{5}\;r({\bm x}_i,{\bm x}_j) +  \frac{5}{3}\;r^2({\bm x}_i, {\bm x}_j)\right )\nonumber \\
&&\times \exp\left ( -\sqrt{5}\;r({\bm x}_i, {\bm x}_j)\right )~~~~
\label{eqn:k_MAT}\\
k_{\rm RQ}({\bm x}_i, {\bm x}_j)  &=& \left ( 1 + \frac{|{\bm x}_i- {\bm x}_j|^2}{2\alpha\ell^2} \right )^{-\alpha}
\label{eqn:k_RQ}
\end{eqnarray}
where $r^2({\bm x}_i, {\bm x}_j) = ({\bm x}_i- {\bm x}_j)^\top \times {M} \times ({\bm x}_i-{\bm x}_j)$ and ${M}$ is a diagonal matrix with different length-scales $\ell_d$ for each dimension of ${\bm x}_i$. The unknown parameters of these functions are found 
by maximizing the log \emph{marginal likelihood} function,

\begin{eqnarray}
\log p(\bm{y}|\bm X,\bm{\theta}) = -\frac{1}{2} \bm{y}^\top K^{-1} \bm{y} - \frac{1}{2}\log |K| -\frac{n}{2} \log (2\pi),
\label{eqn:logml}
\end{eqnarray}
where $\boldsymbol{\theta}$ denotes collectively the parameters of the analytical function for $k(\cdot, \cdot)$ and $|K|$ is the determinant of the matrix $K$. Given the kernel functions thus found, Eq. (\ref{meanGP}) is a GP model, which can be used to make a prediction by interpolation. 

\subsection{Model selection criteria}
\label{subsec:BIC}

As Eq. (\ref{meanGP}) clearly shows, the GP models with different kernel functions will generally have a different predictive power. 
In principle, one could use the marginal likelihood as a metric to compare models with different kernels. However, different kernels have different numbers of free parameters and  
the second term of Eq. (\ref{eqn:logml}) directly depends on the number of parameters in the kernel. This makes the log marginal likelihood undesirable to compare kernels of different 
complexity. 

As shown in Ref. \cite{bic}, a better metric could be the Bayesian information criterion (BIC) defined as
\begin{eqnarray}
\text{BIC}({\cal M}_i) = \log p({\bm y} | {\bm x}, \hat{\mathbf{\theta}}, {\cal M}_i) -\frac{1}{2}|{\cal M}_i|\log n
\label{BIC-eq}
\end{eqnarray}
where $|{\cal M}_i|$ is the number of kernel parameters of the kernel ${\cal M}_i$. In this equation, $ p({\bm y} | {\bm x}, \hat{\mathbf{\theta}}, {\cal M}_i)$ is the marginal likelihood for the optimized kernel $\hat{\boldsymbol{\theta}}$ which maximizes the logarithmic part. The assumption -- one that will be tested in the present work for physics applications -- is that more physical models have a larger BIC. The last term in Eq. (\ref{BIC-eq}) penalizes kernels with a larger number of parameters. The optimal BIC will thus correspond to the kernel yielding the largest value of the log marginal likelihood  function with the fewest number of free parameters.

\section{Physical extrapolation by generalization with Gaussian Processes}
\label{sec:GP_extrapolation}

As shown Refs. \cite{kernel_comb,gp-ss},  one can use the BIC to increase the generalization power of GP models. 
The algorithm proposed in Refs. \cite{kernel_comb,gp-ss} aims to build up the complexity of kernels, starting from the simple kernels (\ref{eqn:k_LIN}) - (\ref{eqn:logml}), in a greedy search algorithm guided by the values of the BIC. Here, we employ this algorithm to extrapolate the quantum properties embodied in lattice models across phase transitions. 

\subsection{Learning with kernel combinations}
\label{sec:gpss_algorithm}

The approach adopted here starts with the simple kernels (\ref{eqn:k_LIN}) - (\ref{eqn:logml}). For each of the kernels, a GP model is constructed and the BIC is calculated. The kernel corresponding to the highest BIC is then selected as the best kernel. We will refer to such kernel as the `base' kernel and denote it by $k_0$. 
The base kernel is  then combined with each of the kernels (\ref{eqn:k_LIN}) - (\ref{eqn:logml}). The new `combined' kernels are chosen to be either of the sum form 
\begin{eqnarray}
c_0 k_0 + c_i k_i
\label{eqn:k_add}
\end{eqnarray}
or of the product form 
\begin{eqnarray}
c_i \times k_0 \times k_i,
\label{eqn:k_mult}
\end{eqnarray}
where $c_0$ and $c_i$ are treated as independent constants to be found by the maximization of the log marginal likelihood. 
The GP models with each of the new kernels are constructed and the BIC values are calculated. The kernel of the model with the largest BIC is then chosen as $k_0$ and the process is iterated. We thus have an `optimal policy' algorithm \cite{RL} that selects the kernel assumed optimal based on the BIC at every step in the search. 

We note that a similar procedure could be used to improve the accuracy of GP models for the interpolation problems. 
We have done this in one of our recent articles \cite{jun-paper}, where GP models were used to construct a six-dimensional potential energy surface for a chemically reactive complex with a very small number of training points.
In the case of interpolation problems, it may also be possible to use cross-validation for kernel selection \cite{arthur-paper}.  Cross-validation could also be applied to kernel selection for the extrapolation problems. 
We have not attempted to do this in the present work. We will compare the relative performance of the validation error and the BIC as the kernel selection metric in a future work. 

\section{Extrapolation of quantum properties}
\label{sec:extrap_quantum_prop}

In this section, we present the results illustrating the performance of the algorithm described above for the prediction of the quantum properties of complex systems outside the range of the training data. Our particular focus is on predicting properties that undergo a sharp variation or discontinuity at certain values of the Hamiltonian parameters. Such properties cannot be directly modelled by GPs because the mean of a GP is a smooth, differentiable function. 

 The main idea \cite{extrapolation-paper} is to train a GP model with functions (obtained from the solutions to the Schr\"{o}dinger equation) that vary smoothly across phase transitions and derive the properties undergoing sharp changes from such smoothly varying function. We thus expect this procedure to be generally applicable to extrapolation across second-order phase transitions. Here, we present two examples to illustrate this. The particular focus of the discussion presented below is on how the method converges to the accurate predictions as the complexity of the kernels increases.  

\subsection{Extrapolation across sharp polaron transitions}
\begin{figure}[h!]
	\includegraphics[width=0.45\columnwidth]{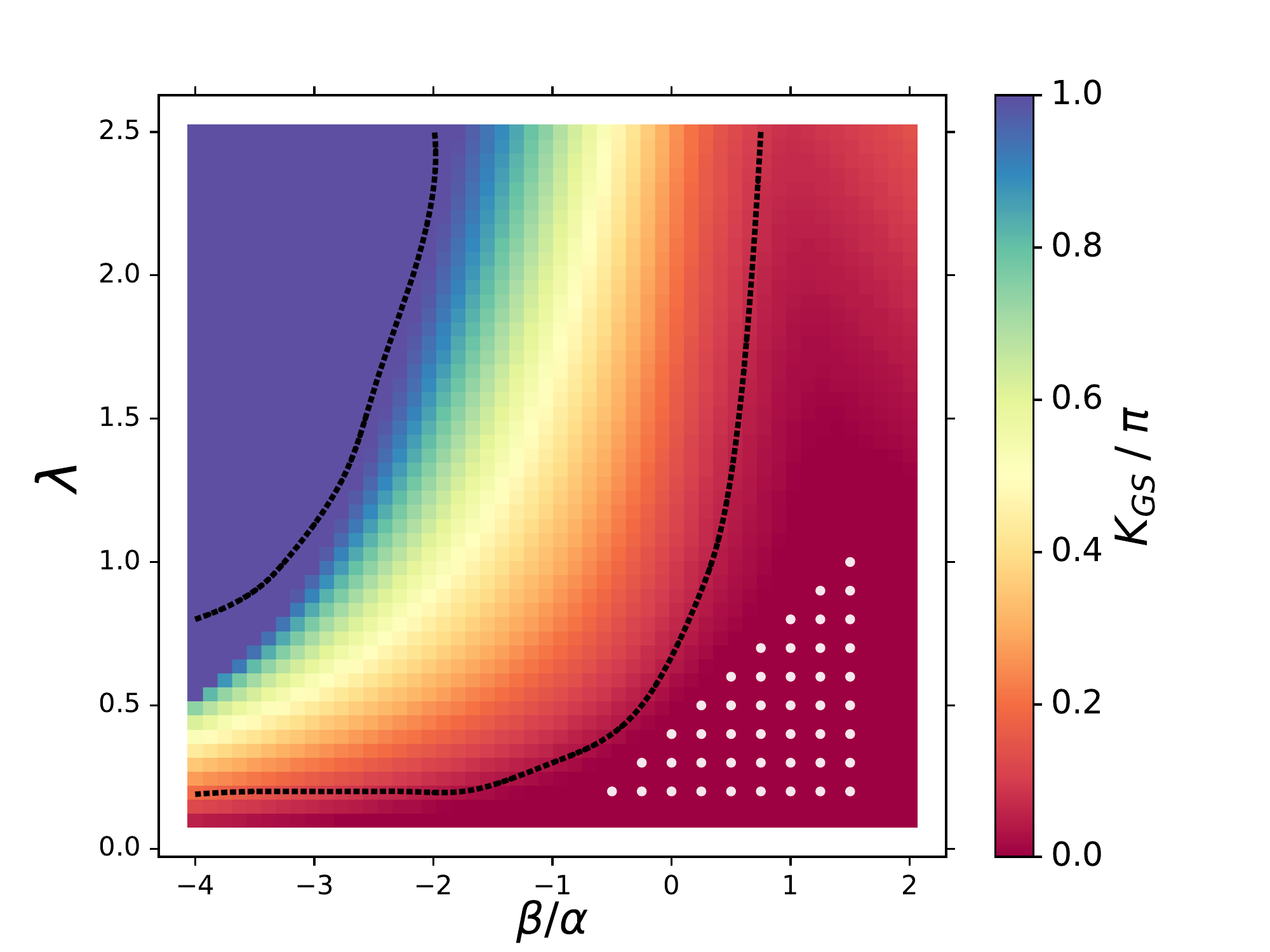}
	\includegraphics[width=0.45\columnwidth]{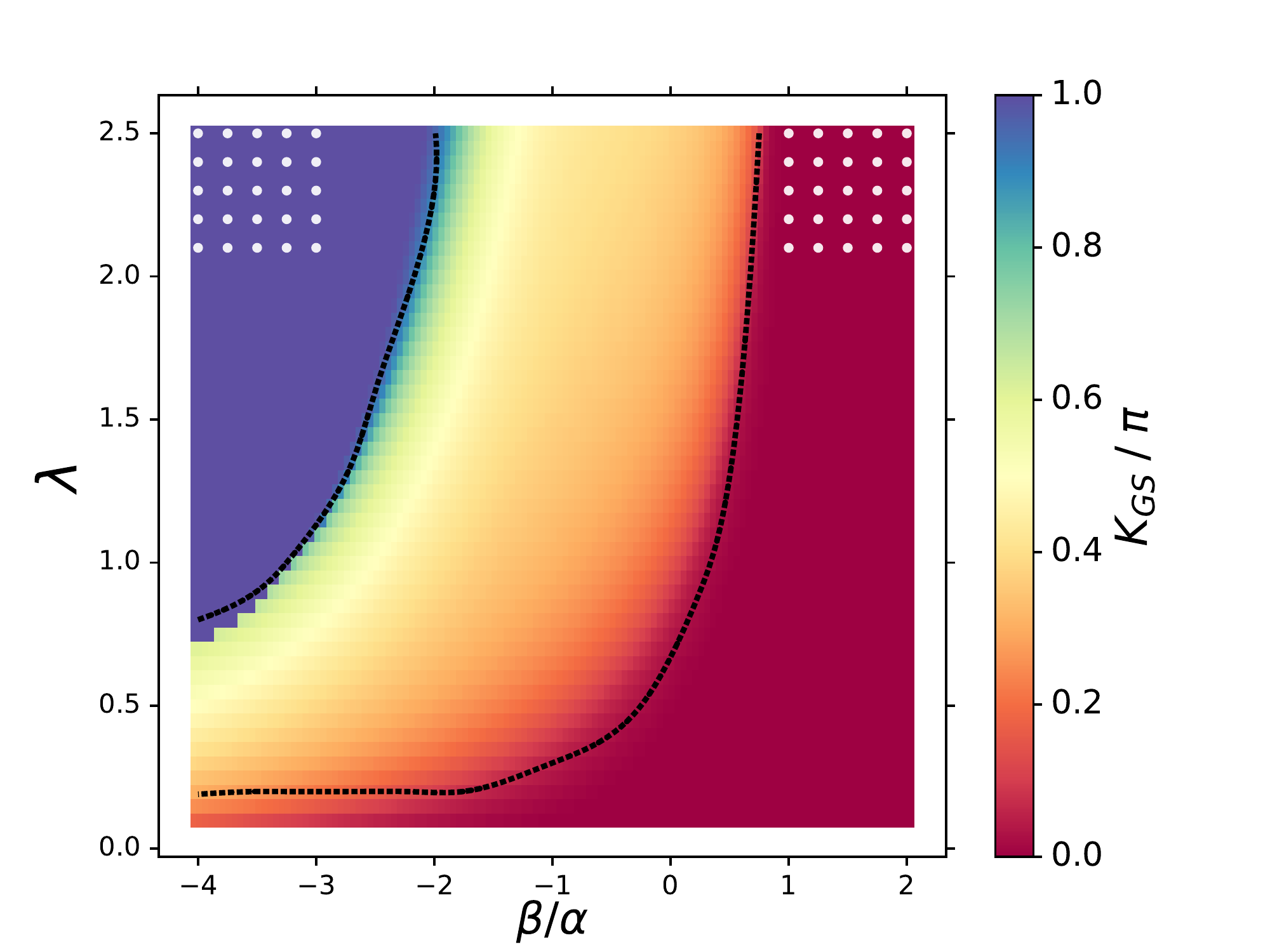}
	\caption{Adapted with permission from Ref. \cite{extrapolation-paper}, Copyright \copyright~APS, 2018. 
	The polaron ground state momentum $K_{GS}$  for the mixed model (\ref{polaron}) as a function of $\beta/\alpha$ for $\lambda = 2\alpha^2/t \hbar\omega$. 
	The color map is the prediction of the GP models. 
	The curves are the quantum calculations from Ref. \cite{HerreraPRL}.  The models are trained by the polaron dispersions at the parameter values indicated by the white dots.
	{{The optimized kernel combination is $(k_{MAT} + k_{RBF})\times k_{LIN}$ (left panel) and $(k_{MAT}\times k_{LIN} + k_{RBF})\times k_{LIN}$ (right panel).}}}
	\label{phases_PRL}
\end{figure}

As discussed in Section \ref{sec:quantum_sys}, the Hamiltonian describing a quantum particle coupled to optical phonons through a combination of two couplings defined by Eq. (2) yields polarons with unusual properties. In particular, it was previously shown \cite{HerreraPRL} that the ground-state momentum of such polarons undergoes two sharp transitions as the ratio $\alpha/\beta$ in Eq. (2) as well as the parameter $\lambda = 2\alpha^2/t \hbar\omega$ are varied. The dimensionless parameter $\lambda$ is defined in terms of the bare particle hopping amplitude $t$ and the phonon frequency $\omega$. It quantifies the strength of coupling between the bare particle and the phonons. One can thus calculate the ground-state momentum or the effective mass of the polaron as a function of $\lambda$ and $\alpha/\beta$. The values of $\lambda$ and $\alpha/\beta$, where the polaron momentum and effective mass undergo sharp changes, separate the `phases' of the Hamiltonian (\ref{polaron}).

The GP models are trained by the {\it polaron energy dispersions} (i.e. the full curves of the dependence of the polaron energy on the polaron momentum) at different values of $\lambda, \alpha$ and $\beta$. These models are then used to generalize the full energy dispersions to values of $\lambda, \alpha$ and $\beta$ outside the range of the training data and the momentum of the polaron with the lowest energy is calculated from these dispersion curves.  The results are shown in Figure \ref{phases_PRL}. Each of the white dots in the phase diagrams depicted specifies the values of $\alpha, \beta$ and $\lambda$, for which the polaron dispersions were calculated and used as the training data. One can thus view the resulting GP models as four-dimensional, i.e. depending on $\alpha$, $\beta$,  $\lambda$ and the polaron momentum. 

Figure \ref{phases_PRL} illustrates two remarkable results: 

\begin{itemize}

\item[$\circ$] The left  panel illustrates that the GP models are capable of predicting {\it multiple} new phase transitions by using the training data {\it entirely} in one single phase. This proves our conjecture \cite{extrapolation-paper} that the evolution of physical properties with the Hamiltonian parameters in a single phase contains information about multiple phases and multiple phase transitions. 

\item[$\circ$]  While perhaps less surprising, the right panel illustrates that the accuracy of the predictions increases significantly and the predictions of the phase transitions become quantitative if the models are trained by data in two phases. The model illustrated in this panel extrapolates the polaron properties from high values of $\lambda$ to low values of $\lambda$.  Thus, the extrapolation becomes much more accurate if the models are trained by data in multiple phases.  

\end{itemize}

\noindent
In the following section we analyze how the kernel selection algorithm described in Ref. \cite{extrapolation-paper} and briefly above arrives at the models used for the predictions in Figure \ref{phases_PRL}.

\subsection{Effect of kernel complexity}
\label{subsec:kernel_complex}

\begin{figure}
\centering
	\includegraphics[width=0.82\columnwidth]{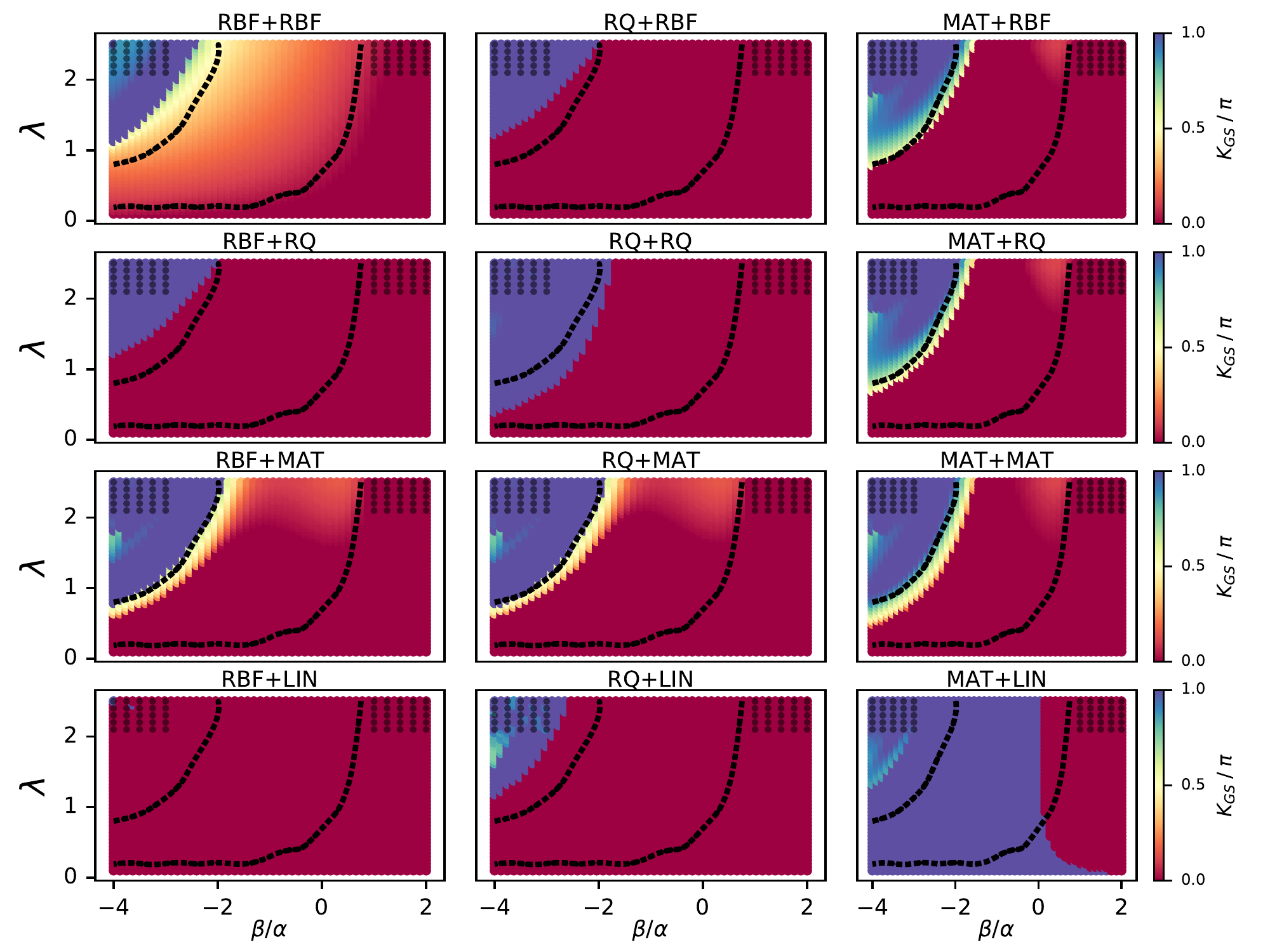}
	\caption{The polaron ground state momentum $K_{GS}$  for the mixed model (\ref{polaron}) as a function of $\beta/\alpha$ for $\lambda = 2\alpha^2/t \hbar\omega$. The black dashed curves are the calculations from Ref. \cite{HerreraPRL}. The color map is the prediction of the GP models with the fully optimized kernels.   The models are trained by the polaron dispersions at the parameter values indicated by the black dots. 
	The different kernels considered here are all possible pairwise additions (\ref{eqn:k_add}) of two simple kernels from the family of kernels ($k_{MAT}$ , $k_{RQ}$ and $k_{RBF}$).}
	\label{phases-again}
\end{figure}

Figure \ref{phases-again} illustrates the performance of the models with kernels represented by a simple addition of two simple kernels, when trained by the data in two phases, as in the rigth panel of Figure \ref{phases_PRL}. The examination of this figure shows that the generalization accuracy, including the prediction of the number of the phase transitions, is sensitive to the kernel combination.  For example, the models with the combination of the RBF and LIN kernels do not predict any phase transitions. Most of the other kernel combinations predict only one of the two transitions. Remarkably, the combination of two RBF kernels already leads to the appearance of the second phase transition, and allows the model to predict the location of the first transition quite accurately. The combination of Figures \ref{phases_PRL} and \ref{phases-again} thus illustrates that the BIC is a meaningful metric to guide the kernel selection algorithm, as it rules out many of the kernels leading to incorrect phase diagrams shown in Figure \ref{phases-again}. The results in Figure \ref{phases-again} also raise the question, how many combinations are required for kernels to allow quantitative predictions?

\begin{figure}
\centering
  \includegraphics[width=.4\textwidth]{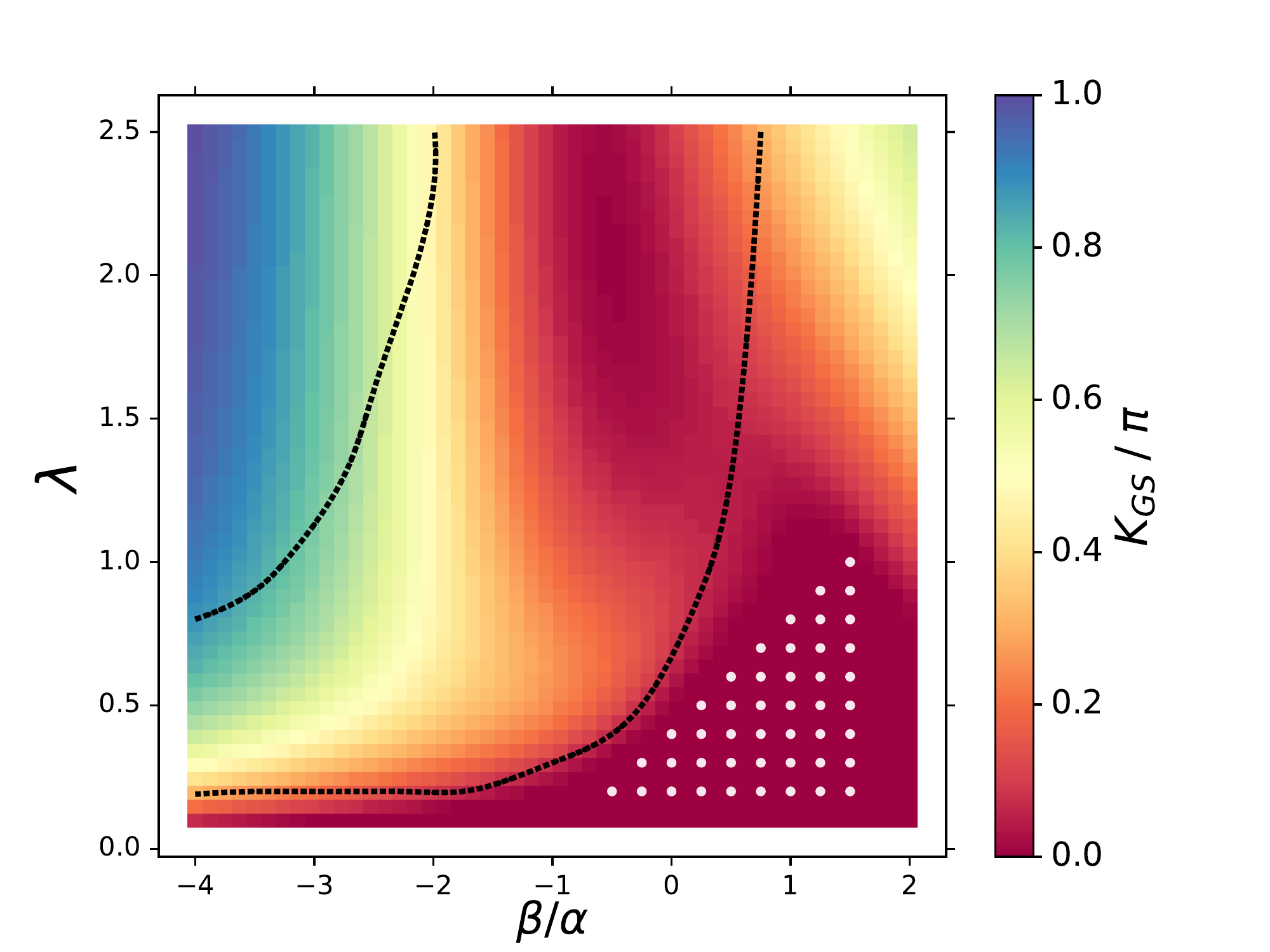}
    \includegraphics[width=.4\textwidth]{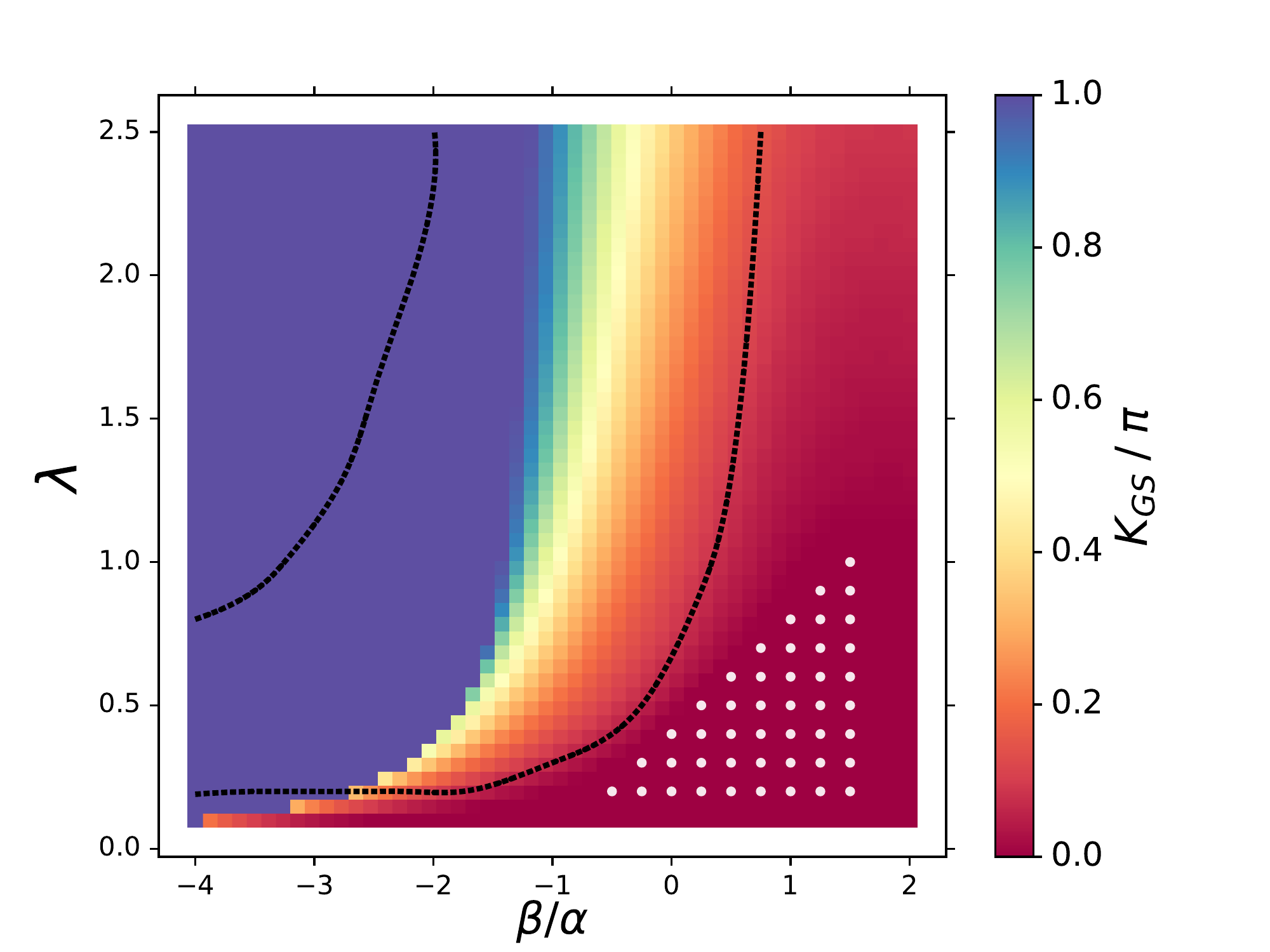}
      \includegraphics[width=.4\textwidth]{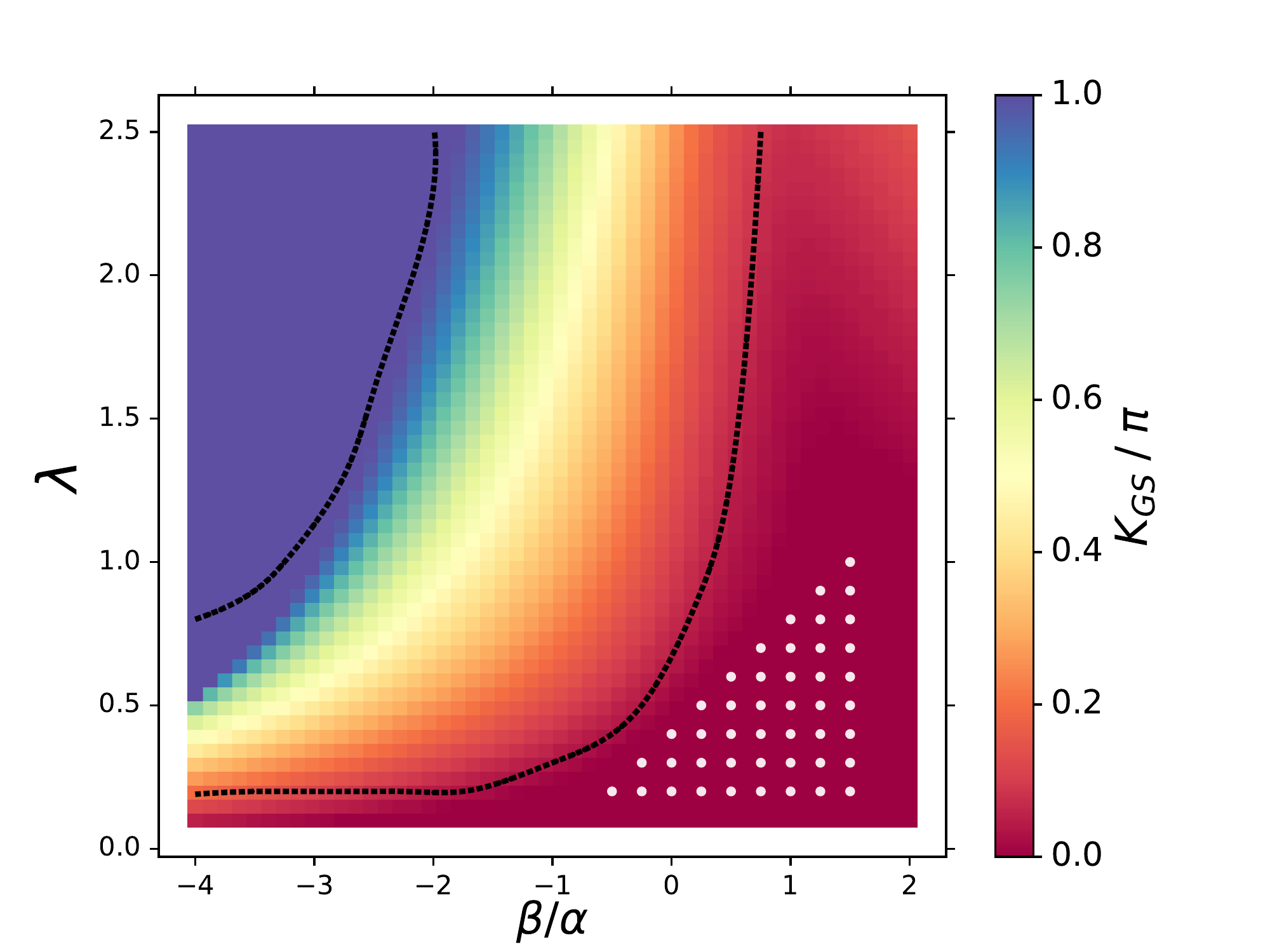}
\caption{
Adapted from the supplementary material of Ref. \cite{extrapolation-paper}. 
Improvement of the phase diagram shown in Figure \ref{phases_PRL} (upper panel) with the kernel complexity increasing as determined by the algorithm described in Section \ref{sec:gpss_algorithm}. 
The panels correspond to the optimized kernels GPL-0 (upper left), GPL-1 (upper right), GPL-2 (lowest panel), where 
 ``GPL-$X$'' denotes the optimal kernel obtained after $X$ depth levels. 
}
\label{phase-diagram_a}
\end{figure}

\begin{figure}[h!]
\centering
  \includegraphics[width=.435\textwidth]{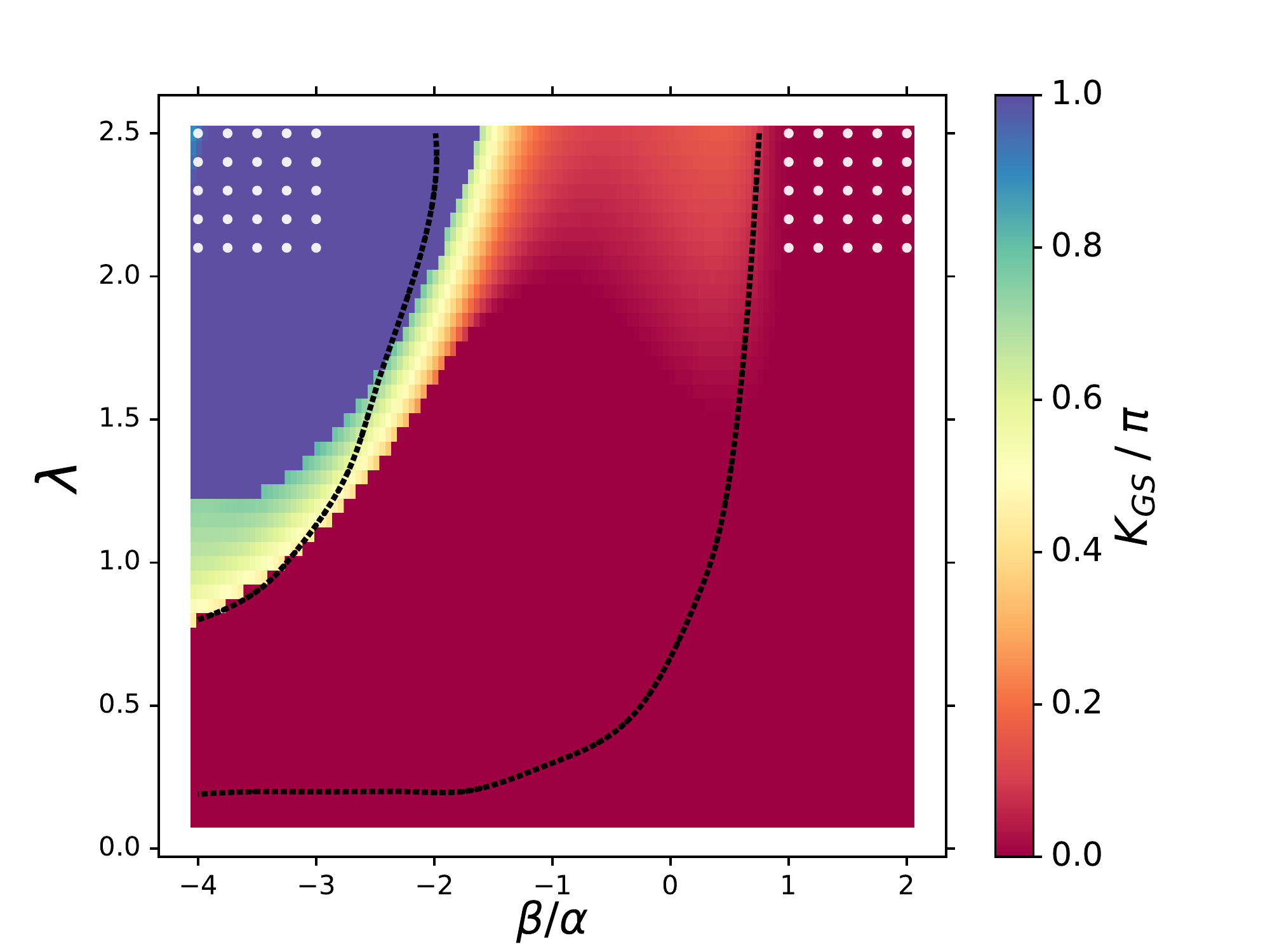}  
  \includegraphics[width=.435\textwidth]{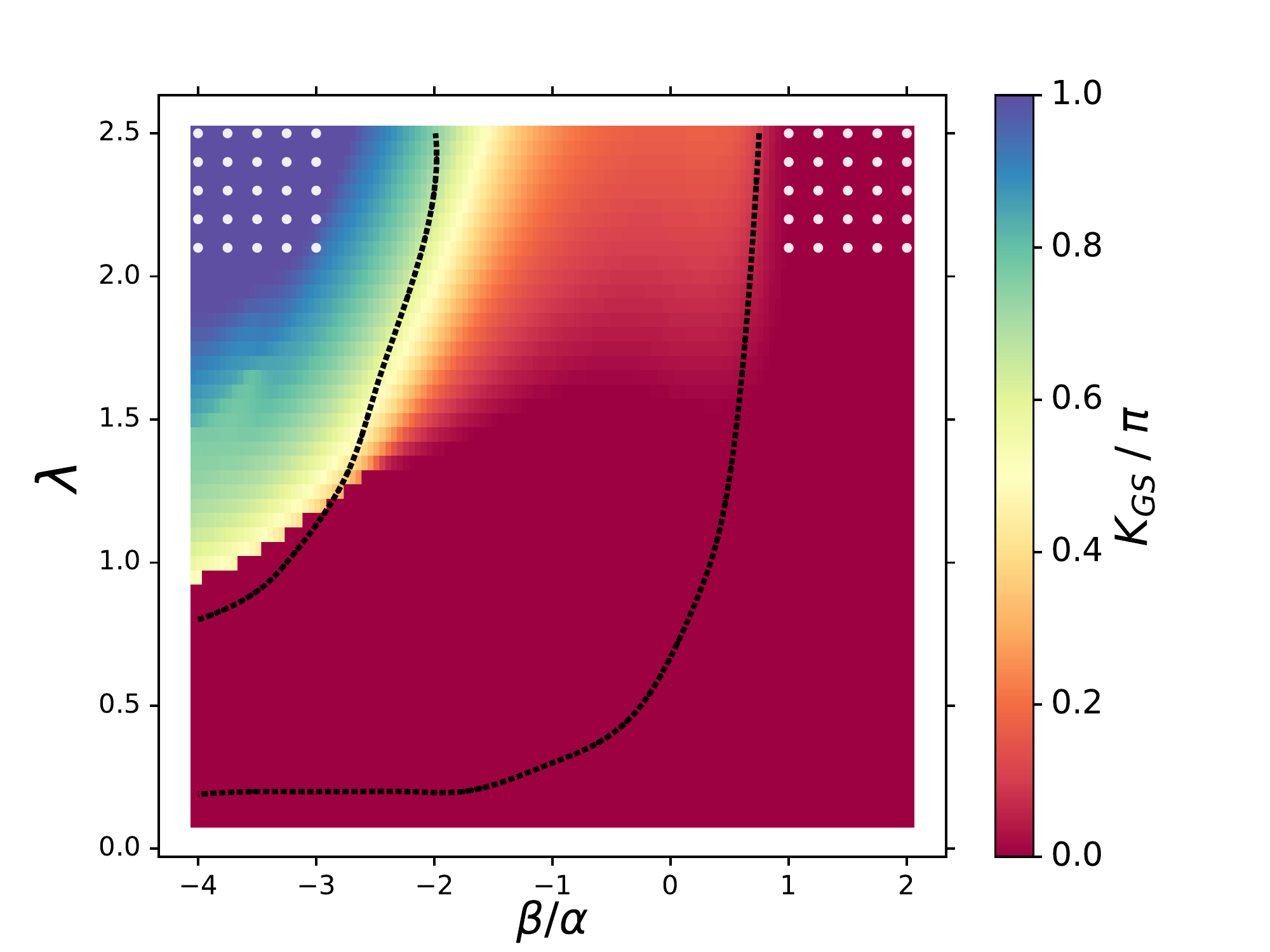}  \\
  \includegraphics[width=.435\textwidth]{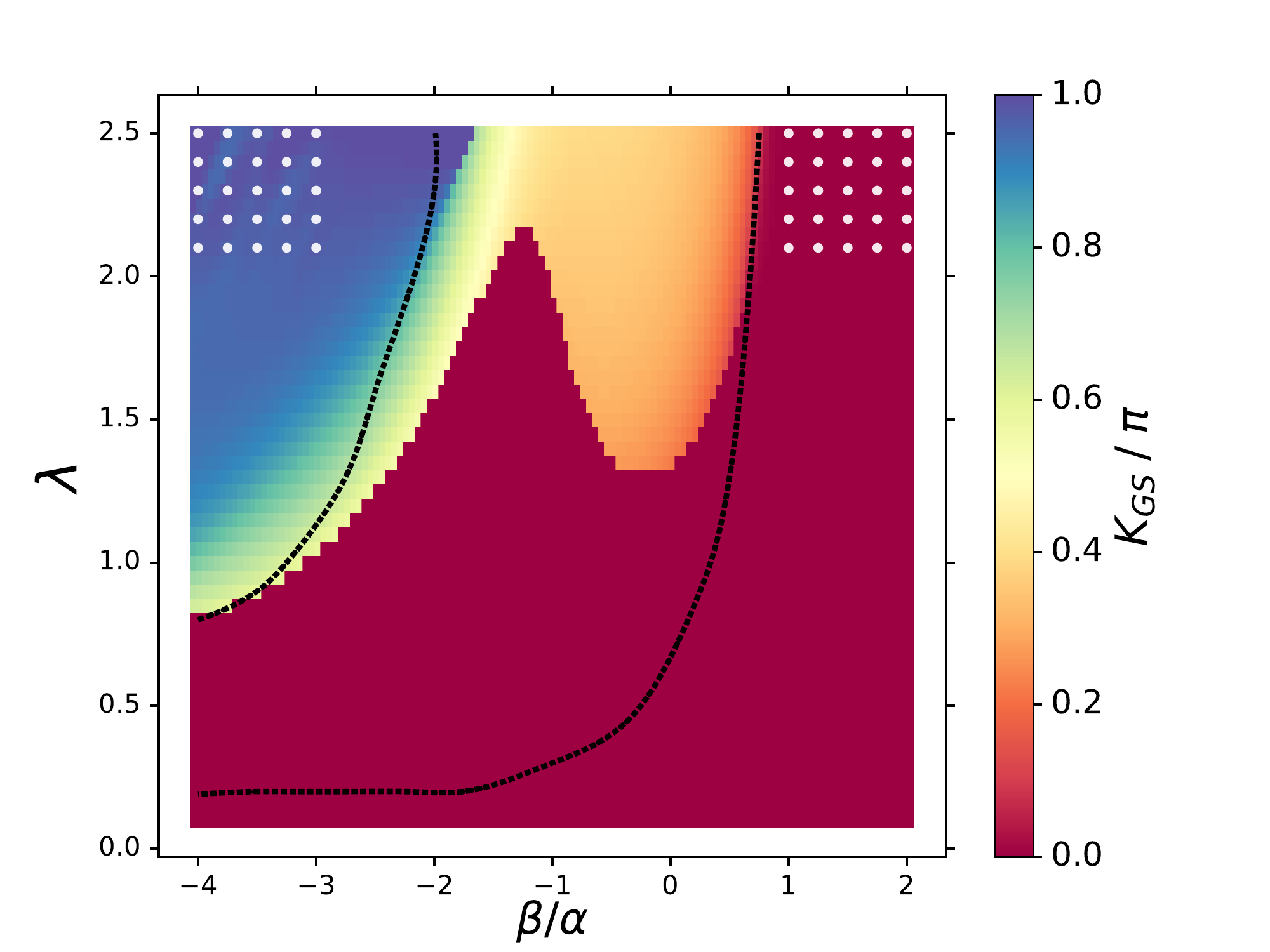}  
  \includegraphics[width=.435\textwidth]{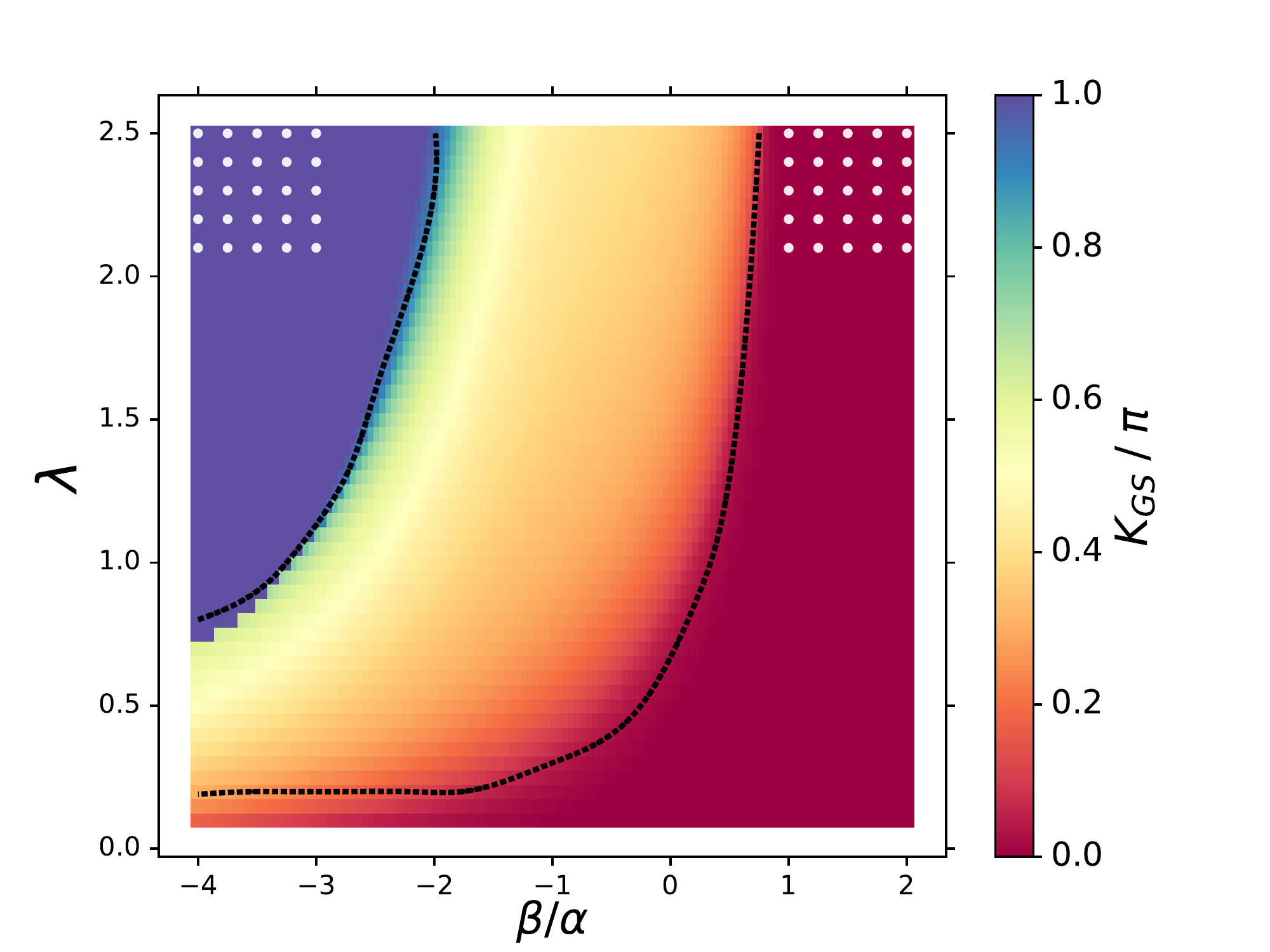}
\caption{
Adapted from the supplementary material of Ref. \cite{extrapolation-paper}. 
Improvement of the phase diagram shown in Figure \ref{phases_PRL} (lower panel) with the kernel complexity increasing as determined by the algorithm depicted in Section \ref{sec:gpss_algorithm}. 
The panels correspond to the optimized kernels GPL-0 (upper left), GPL-1 (upper right), GPL-2 (lower left), GPL-3 (lower right), where 
 ``GPL-$X$'' denotes the optimal kernel obtained after $X$ depth levels. 
}
\label{phase-diagram_b}
\end{figure}

\begin{figure}[h!]
\centering
  \includegraphics[width=.435\textwidth]{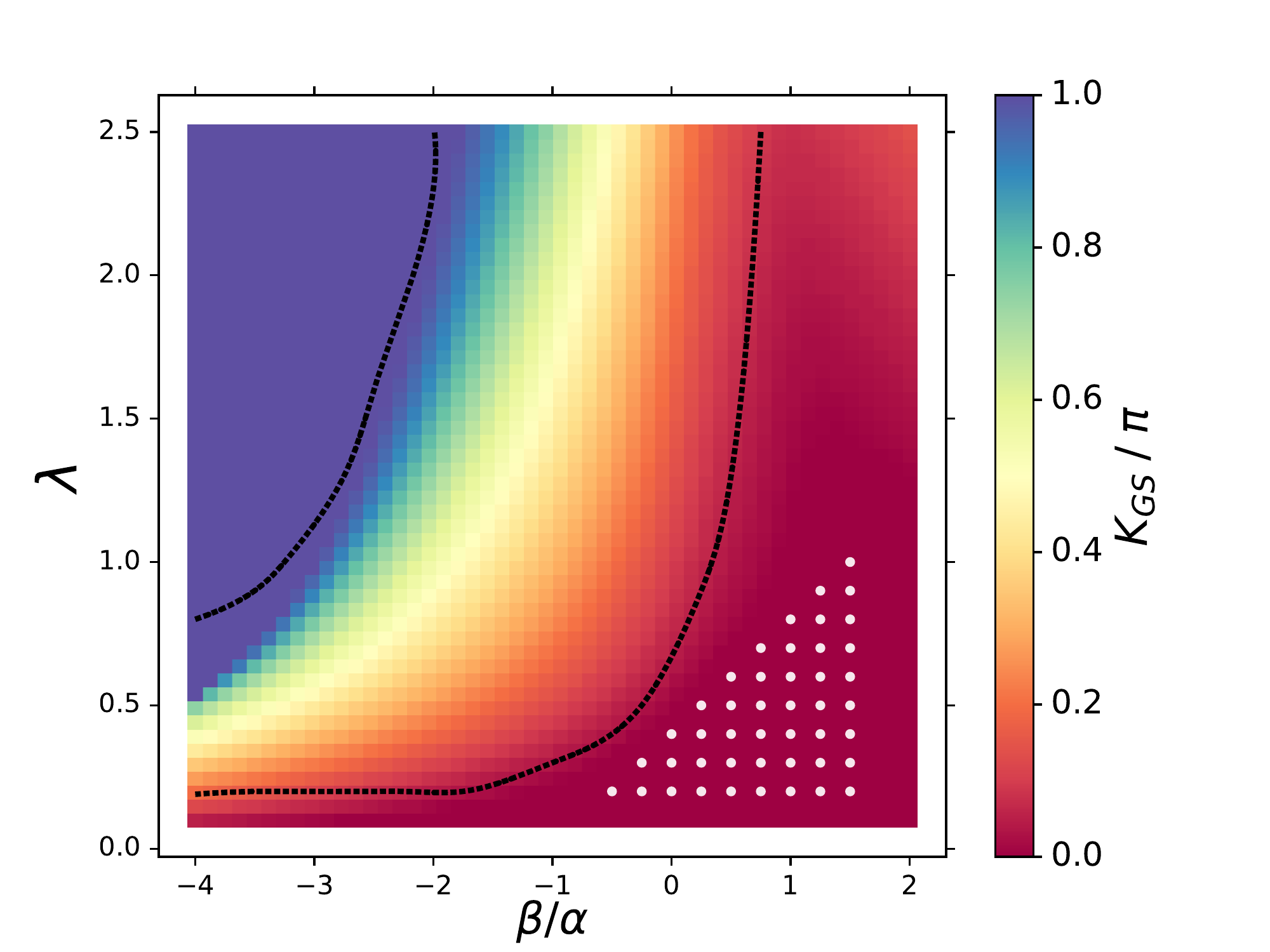}  
  \includegraphics[width=.435\textwidth]{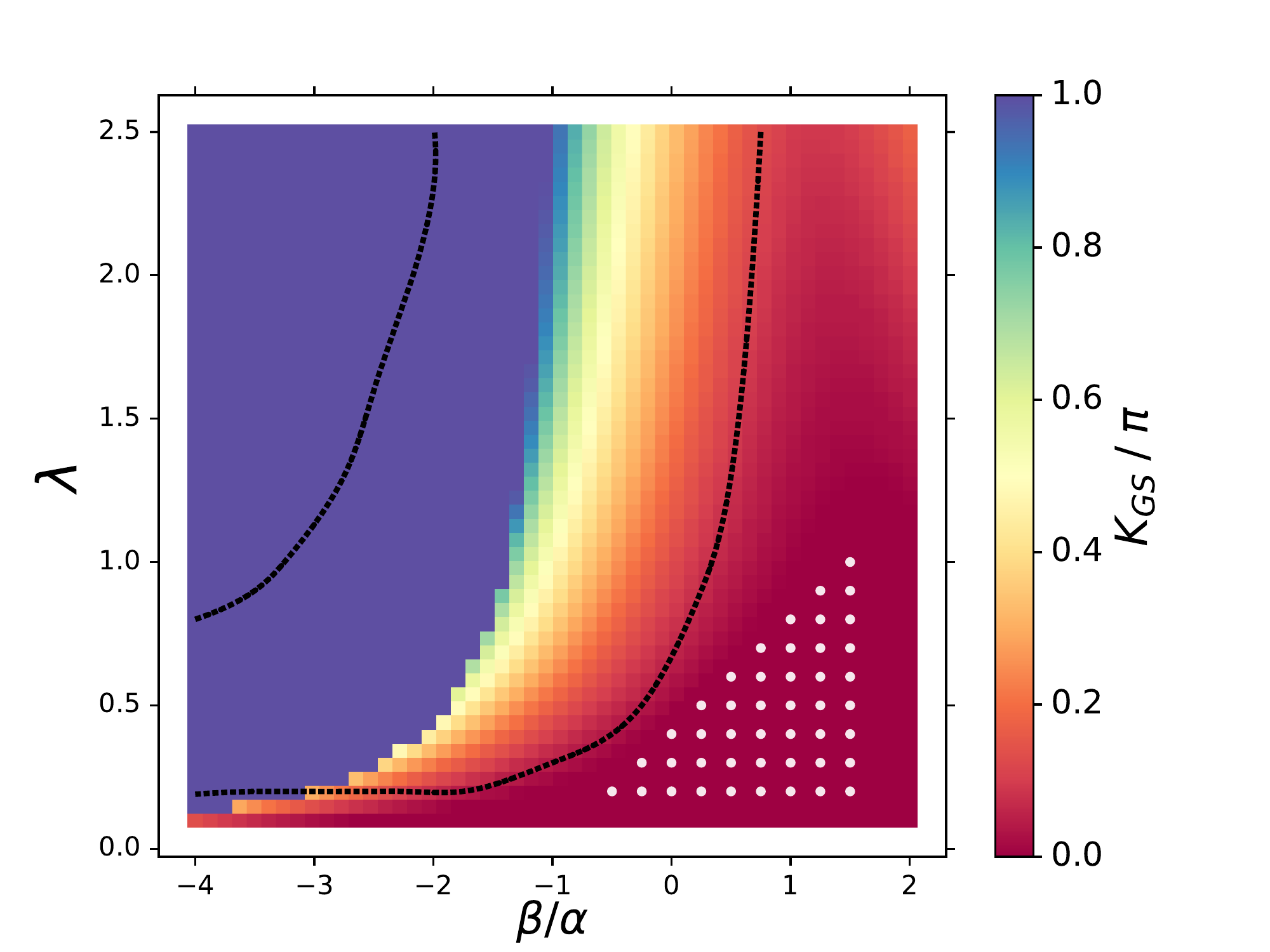}  
  \includegraphics[width=.435\textwidth]{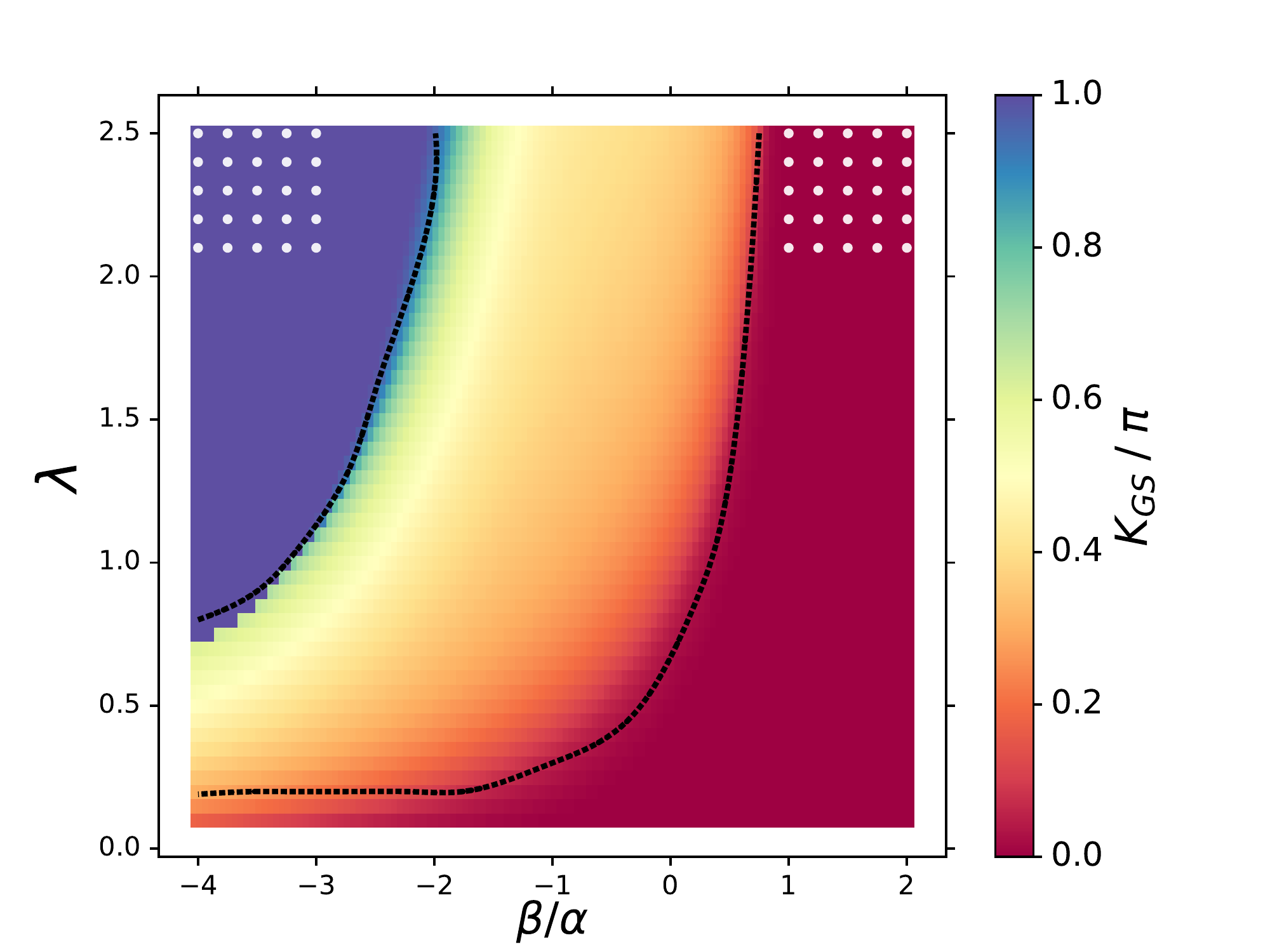}  
  \includegraphics[width=.435\textwidth]{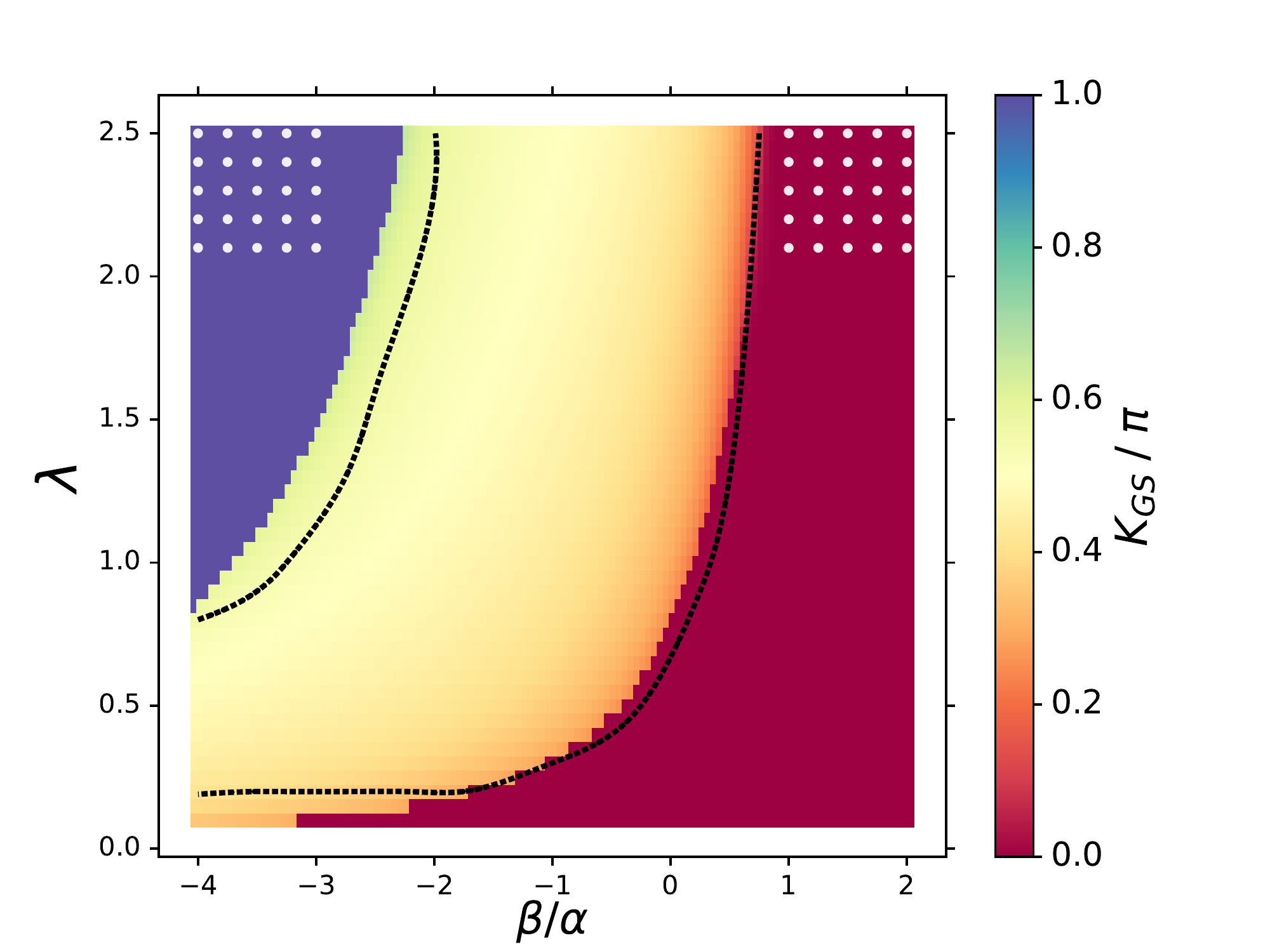}
\caption{Decrease of the prediction accuracy with increasing kernel complexity. Upper panels: left -- GPL-2 (same as the right panel of Figure \ref{phase-diagram_a}),  right -- GPL-3. 
Lower panels: left -- GPL-3 (same as the lower right panel of Figure \ref{phase-diagram_b}), right -- GPL-4. 
}
\label{phase-diagram-decrease}
\end{figure}

To answer this question, we show in Figures \ref{phase-diagram_a} and \ref{phase-diagram_b} the convergence of the phase diagrams to the results in Figure \ref{phases_PRL} with the number of iterations in the kernel selection algorithm. We use the following notation to label the figure panels: GPL-$X$, where $X$ is the number of iteration. Thus, $X = 0$ corresponds to level zero of the kernel selection algorithm, i.e. GPL-0 is the phase diagram predicted by the model with the single simple kernel leading to the highest BIC.  Level $X=1$ corresponds to kernels constructed as the simple combinations (\ref{eqn:k_add}) or (\ref{eqn:k_mult}). Level $X=2$ corresponds to kernels of the form (\ref{eqn:k_add}) or (\ref{eqn:k_mult}), where $k_i$ is a combination of two kernels. As can be seen from Figures \ref{phase-diagram_a} and \ref{phase-diagram_b}, level $X=2$ and $X=3$ produce kernels with sufficient complexity for accurate predictions.

It must be noted that increasing the complexity of the kernels further (by increasing $X$) often decreases the accuracy of the predictions. 
This is illustrated in Figure \ref{phase-diagram-decrease}. 
We assume that this happens either due to overfitting or because the kernels become so complex that it is difficult to optimize them and the maximization of the log marginal likelihood gets stuck in a local maximum. To overcome this problem, one needs to optimize kernels multiple times starting from different conditions (either different sets of training data or different initial kernel parameters) and stop increasing the complexity of kernels when the optimization produces widely different results. Alternatively, the models could be validated by a part of the training data and the complexity of the kernels must be stopped at level $X$ that corresponds to the minimal validation error, as often done to prevent overfitting with NNs \cite{overfitting_NN}. \\

\clearpage
\subsection{Extrapolation across paramagnetic - ferromagnetic transition}
\label{subsec:extrap_Heis}

In this section, we discuss the Heisenberg spin model described by the lattice Hamiltonian 
\begin{eqnarray}
{\cal H} = -\frac{J}{2} \sum_{\langle i,j \rangle } \bar{S}_i \cdot \bar{S}_j ,
\end{eqnarray}
where $\langle i,j \rangle$ only account for nearest-neighbour interactions between different spins $\bar{S}_i$.
The free energy of the system can be calculated within the mean-field approximation to yield 
\begin{eqnarray}
f(T,m) \approx \frac{1}{2}\left (1- \frac{T_c}{T} \right) m^2 + \frac{1}{12}\left (\frac{T_c}{T} \right)^3 m^4,
\label{mf_H}
\end{eqnarray}
where $m$ is the magnetization and $T_c = 1.25$ is the critical temperature of the phase transition between the paramagnetic (${T} > T_c$) and ferromagnetic (${T} < T_c$) phase.\\

\begin{figure}[h!]
\centering
	\includegraphics[width=0.55\columnwidth]{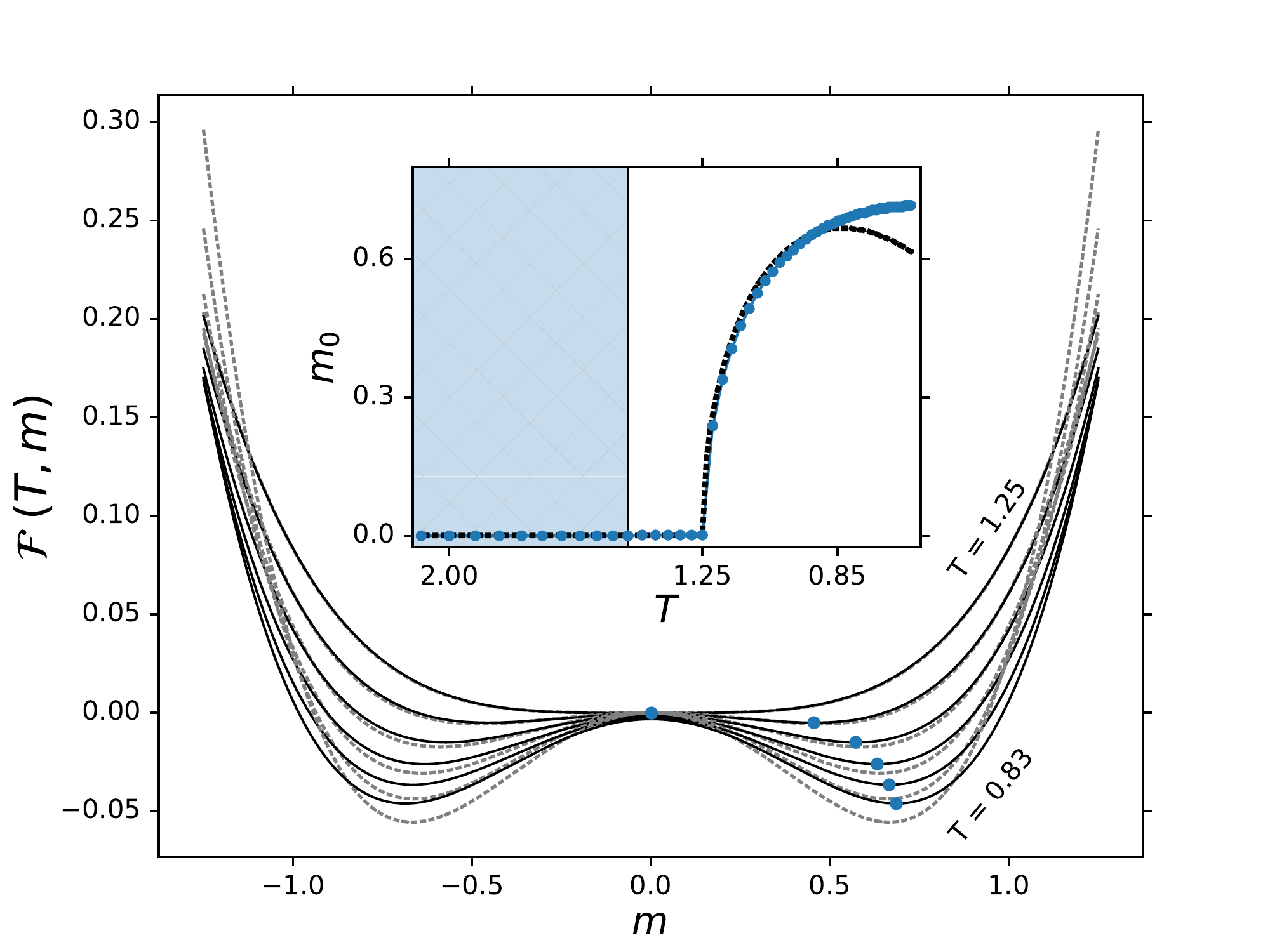}
	\caption{Adapted with permission from Ref. \cite{extrapolation-paper}, Copyright \copyright~APS, 2018. 
	{{GP prediction (solid curves) of the free energy density $f(T,m)$ of the mean-field Heisenberg model produced by Eq.  (\ref{mf_H}) (dashed curves).
	Inset: the order parameter $m_0$ that minimizes $f(T,m)$: symbols -- GP predictions, dashed curve -- from Eq. (\ref{mf_H}). 
	The GP models are trained with 330 points at $1.47 <T < 2.08$ (shaded area) and $-1.25 <m < 1.25$. 		}}}
\label{heisenberg}
\end{figure}

We train GP models by the entire free-energy curves  at temperatures far above $T_c$. The free energy curves are then predicted by the extrapolation models 
at temperatures decreasing to the other side of the transition. The oder parameter $m_0$ -- defined as the value of magnetization that minimizes free energy -- is then computed from the extrapolated predictions. The results are shown in Figure \ref{heisenberg}. 

\begin{figure}[h!]
\centering
  \includegraphics[width=.4\textwidth]{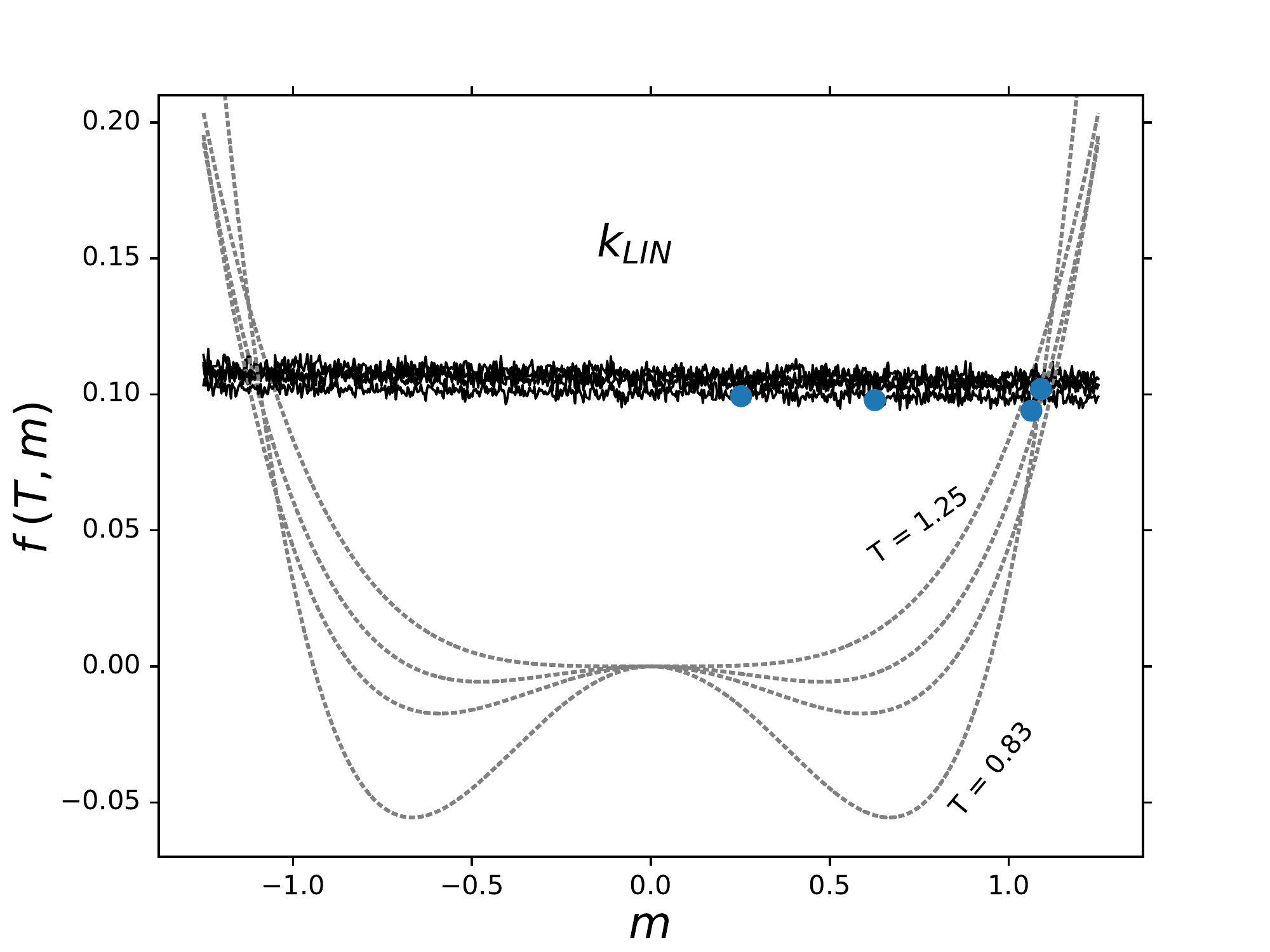} 
  \includegraphics[width=.4\textwidth]{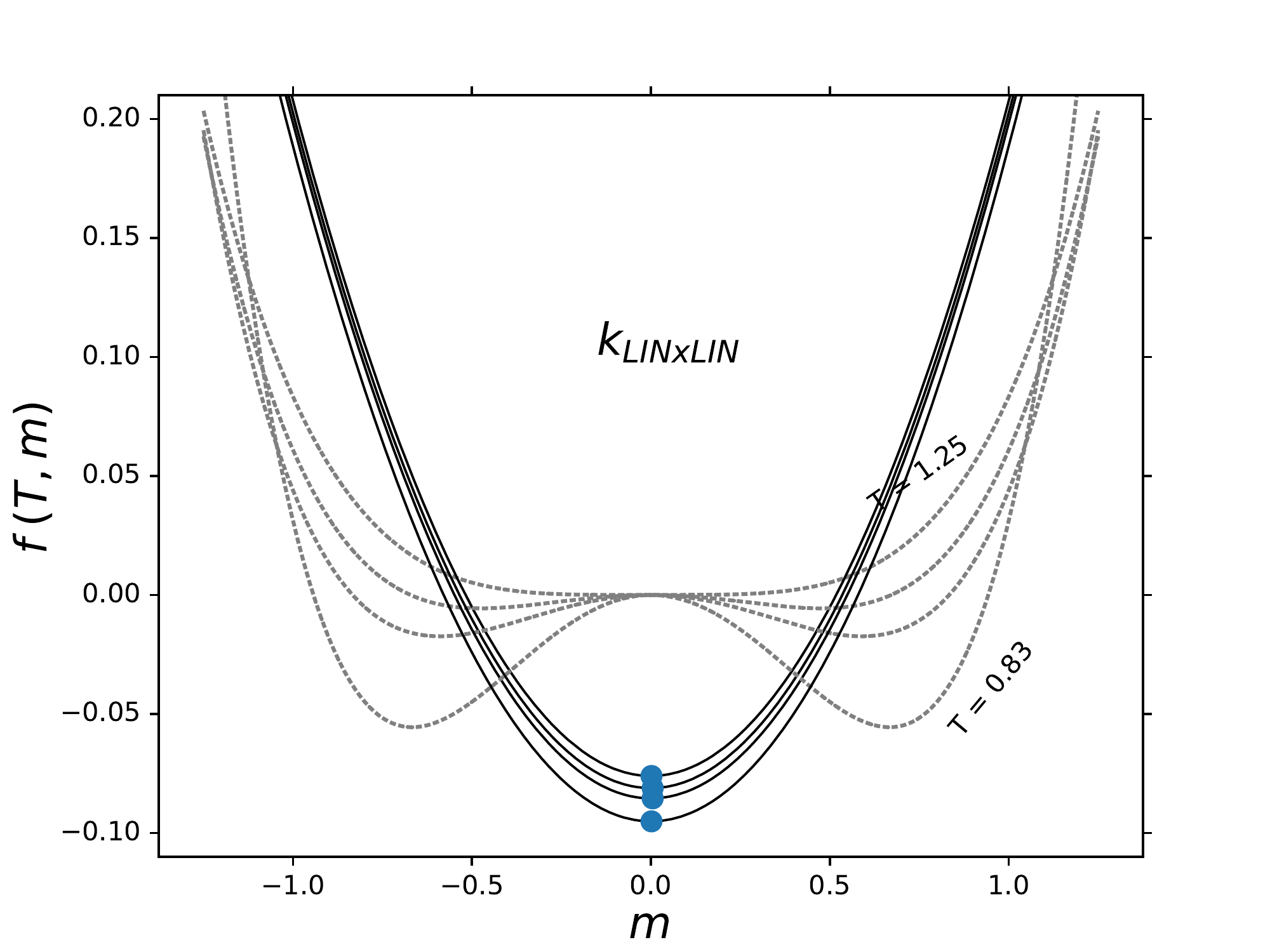} 
  \includegraphics[width=.4\textwidth]{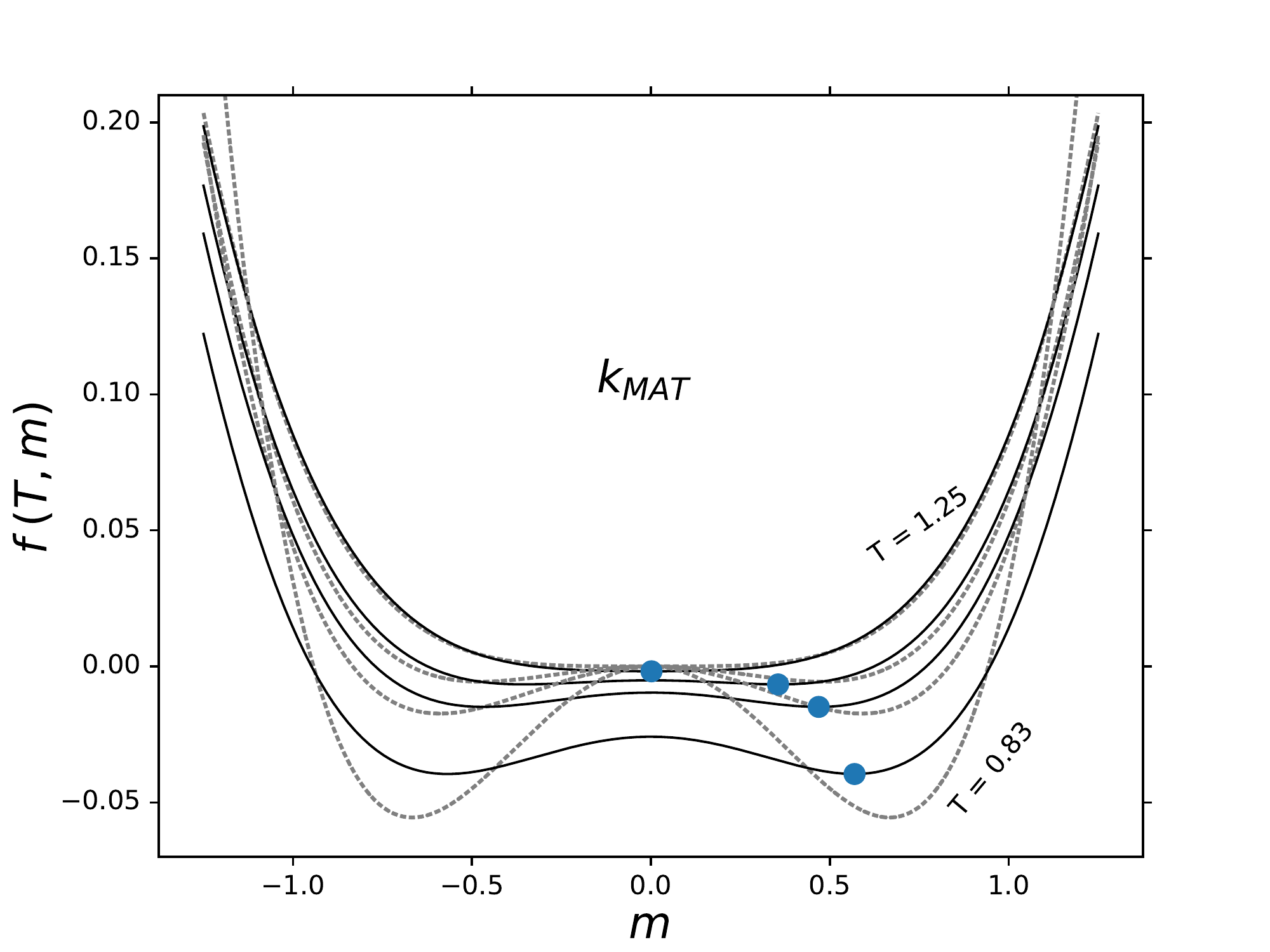} 
  \includegraphics[width=.4\textwidth]{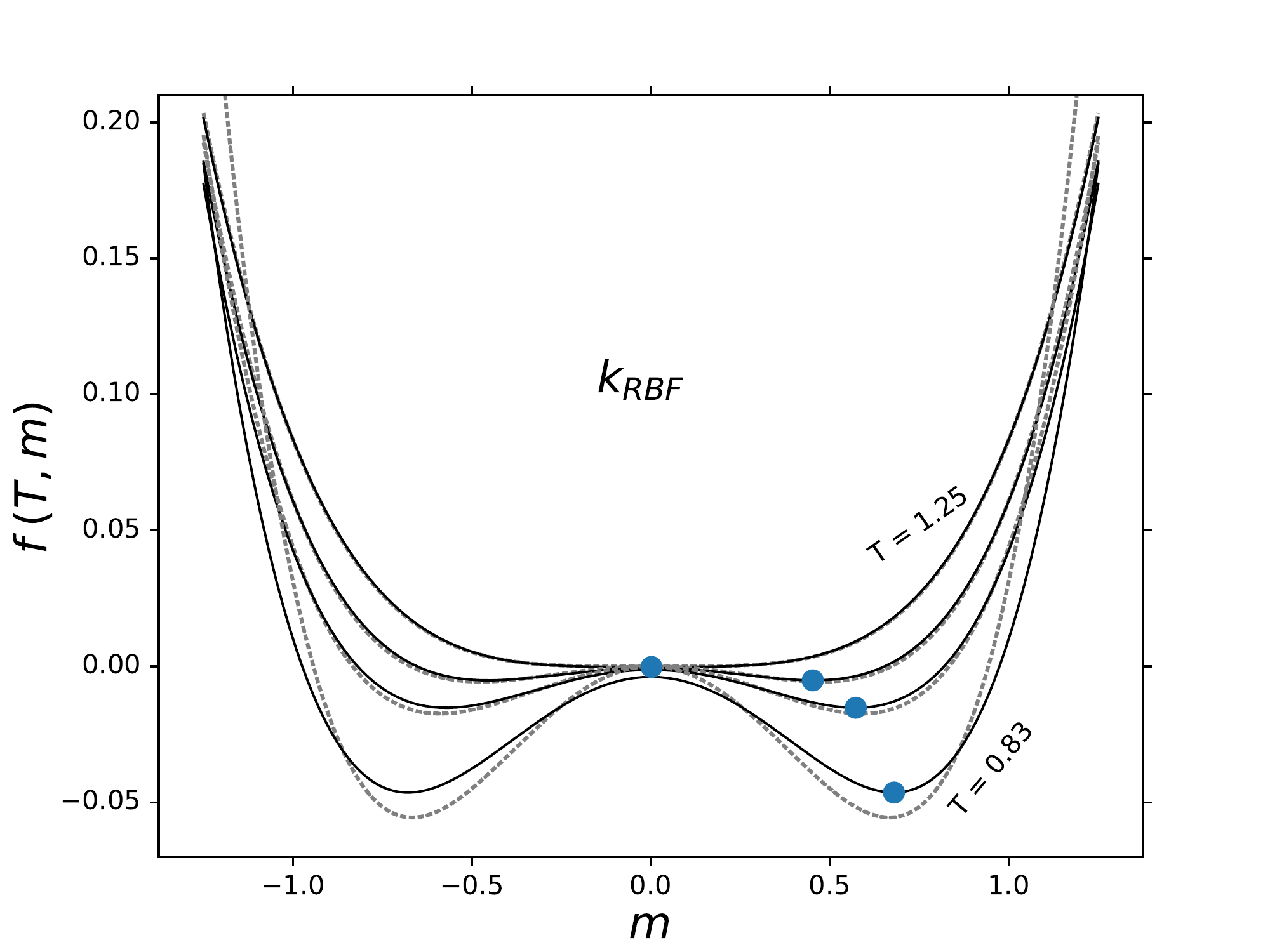}     
\caption{GP prediction (solid curves) of the free energy density $f(T,m)$ of the mean-field Heisenberg model produced by Eq. (18) (dashed curves). All GP models are trained with 330 points at $1.47 < T < 2.08$ (shaded area) and $-1.25 < m < 1.25$. The kernel function used in the GP models is indicated in each panel.}
\label{single-kernel-bad-or-good}
\end{figure}

\begin{figure}
\centering
  \includegraphics[width=.6\textwidth]{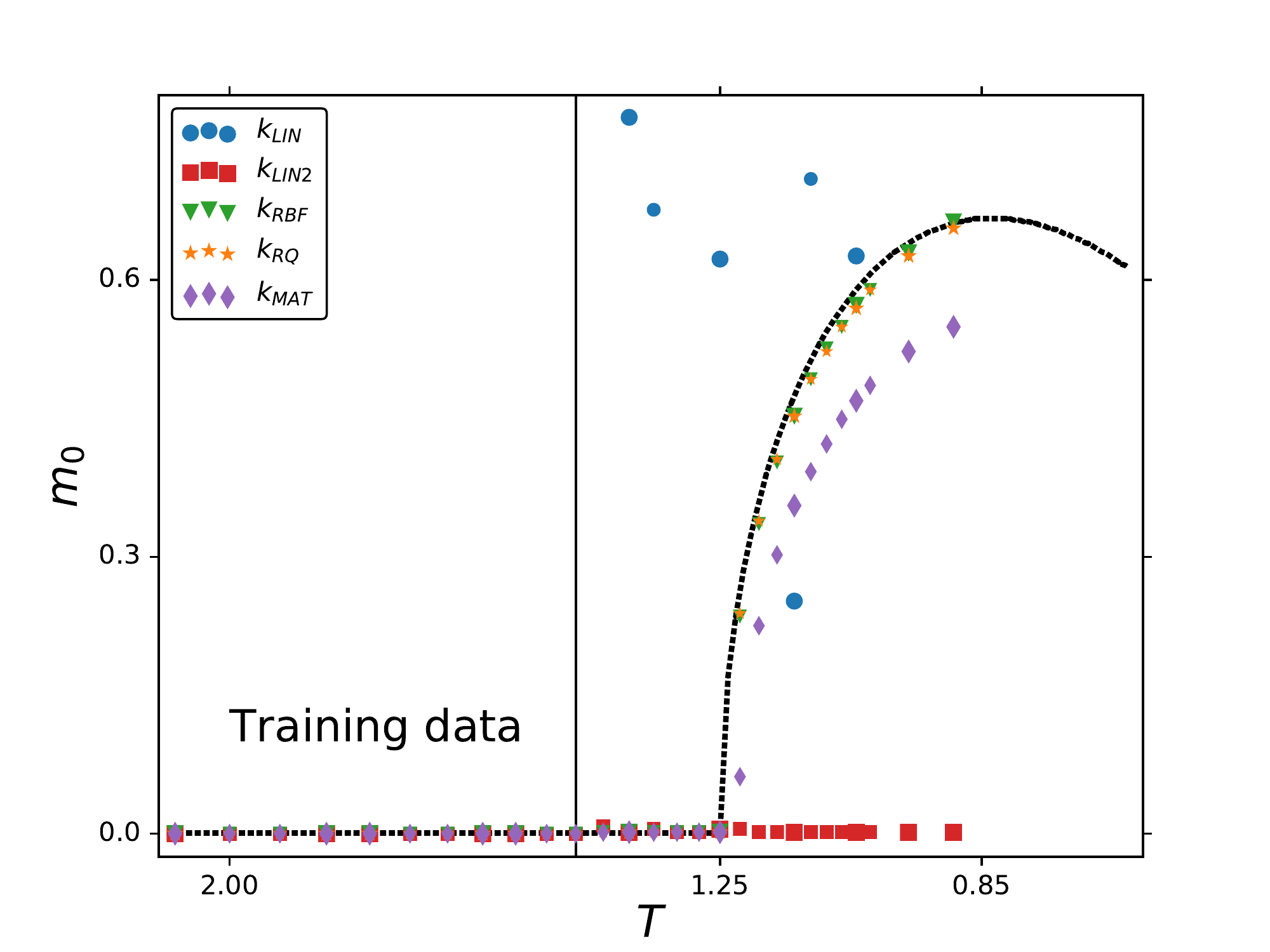}  
\caption{The order parameter $m_0$ that minimizes $f(T,m)$: symbols -- GP predictions, dashed curve -- from Eq. (\ref{mf_H}). The order parameter $m_0$ is computed with the GP predictions using different kernels, illustrated in Figure (\ref{single-kernel-bad-or-good}). }
\label{single-kernel-bad-or-good-order-parameter}
\end{figure}

As evident from Eq. (\ref{mf_H}), the free-energy curves have an analytic dependence on temperature $T$ so this is a particularly interesting case for testing the generalization models. 
Can the kernel selection algorithm adopted here converge to a model that will describe accurately the analytic dependence of the free energy (\ref{mf_H}) as well as the order parameter derived from it? We find that the temperature dependence of Eq. (\ref{mf_H}) can be rather well captured and accurately generalized by a model already with one simple kernel! However, this kernel must be carefully selected. As Figure \ref{single-kernel-bad-or-good} illustrates, the accuracy of the free-energy prediction varies widely with the kernel. This translates directly into the accuracy of the order-parameter prediction illustrated by Figure \ref{single-kernel-bad-or-good-order-parameter}. Figure \ref{single-kernel-bad-or-good-order-parameter} illustrates that the RBF and RQ kernels 
capture the evolution of the order parameter quantitatively, while the LIN, MAT and quadratic kernels produce incorrect results. 

Table \ref{table-1} lists the BIC values for the models used to obtained the results depicted in Figure \ref{single-kernel-bad-or-good-order-parameter}, clearly demonstrating that the higher value of the BIC corresponds to the model with the better prediction power. 

\begin{table}
\centering
\begin{tabular}{|c | c|}
\hline \hline
Kernel type & BIC \\ \hline
RQ & 8667.10 \\
RBF & 8657.13 \\
MAT & 7635.20 \\
LIN & - 104437128213.0 \\
LIN$\times$LIN & -10397873744.9 \\ \hline \hline
\end{tabular}
\caption{The numerical values of the BIC (\ref{BIC-eq}) for the models with different simple kernels (\ref{eqn:k_LIN}) - (\ref{eqn:logml}) used for the predictions of the order parameter depicted in Figure \ref{single-kernel-bad-or-good-order-parameter}.}
\label{table-1}
\end{table}

\subsection{Validation of extrapolation}
\label{subsec:val_extrap_pred}

\begin{figure}[h!]
\centering
  \includegraphics[width=.45\textwidth]{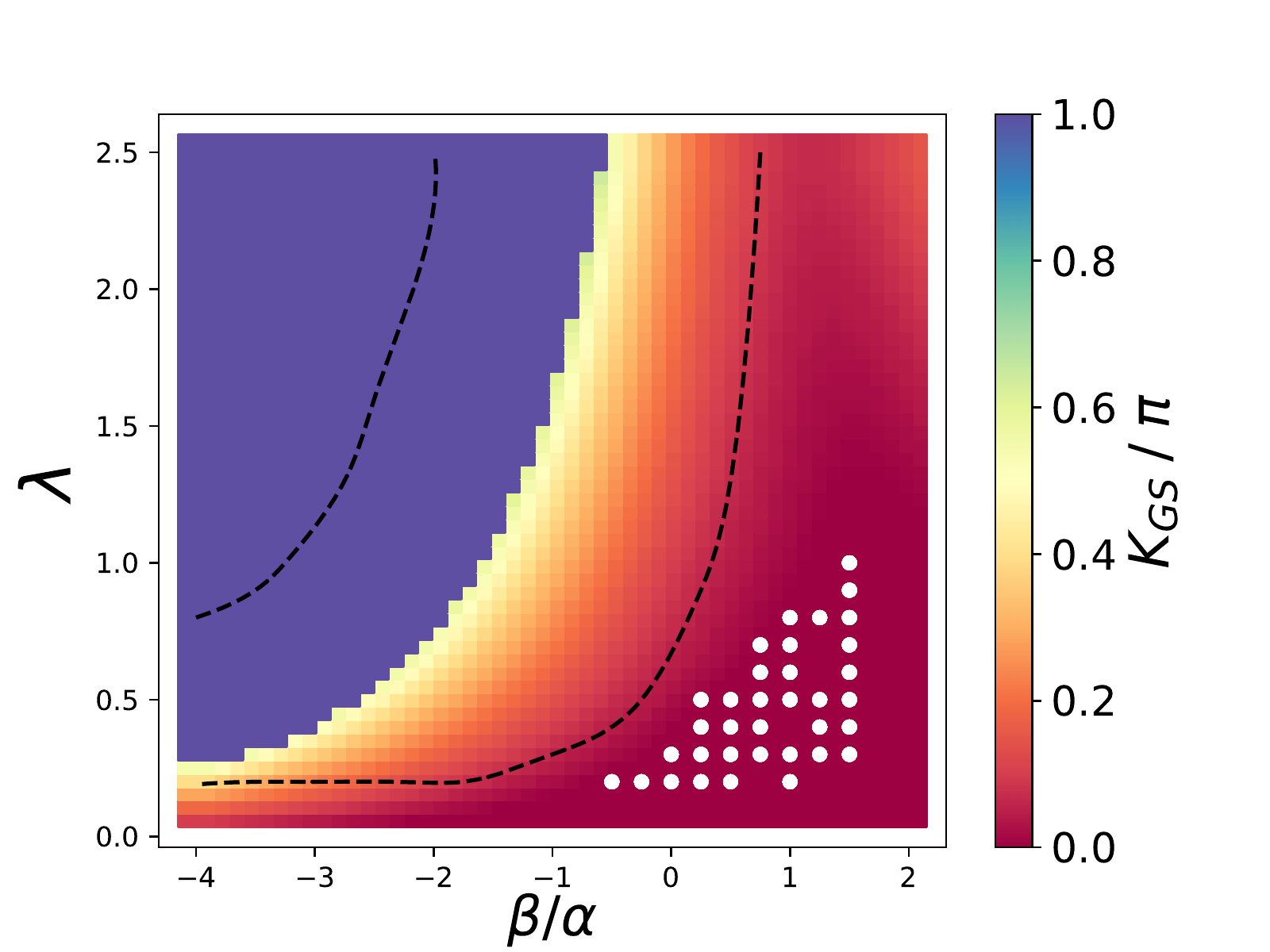}  
  \includegraphics[width=.45\textwidth]{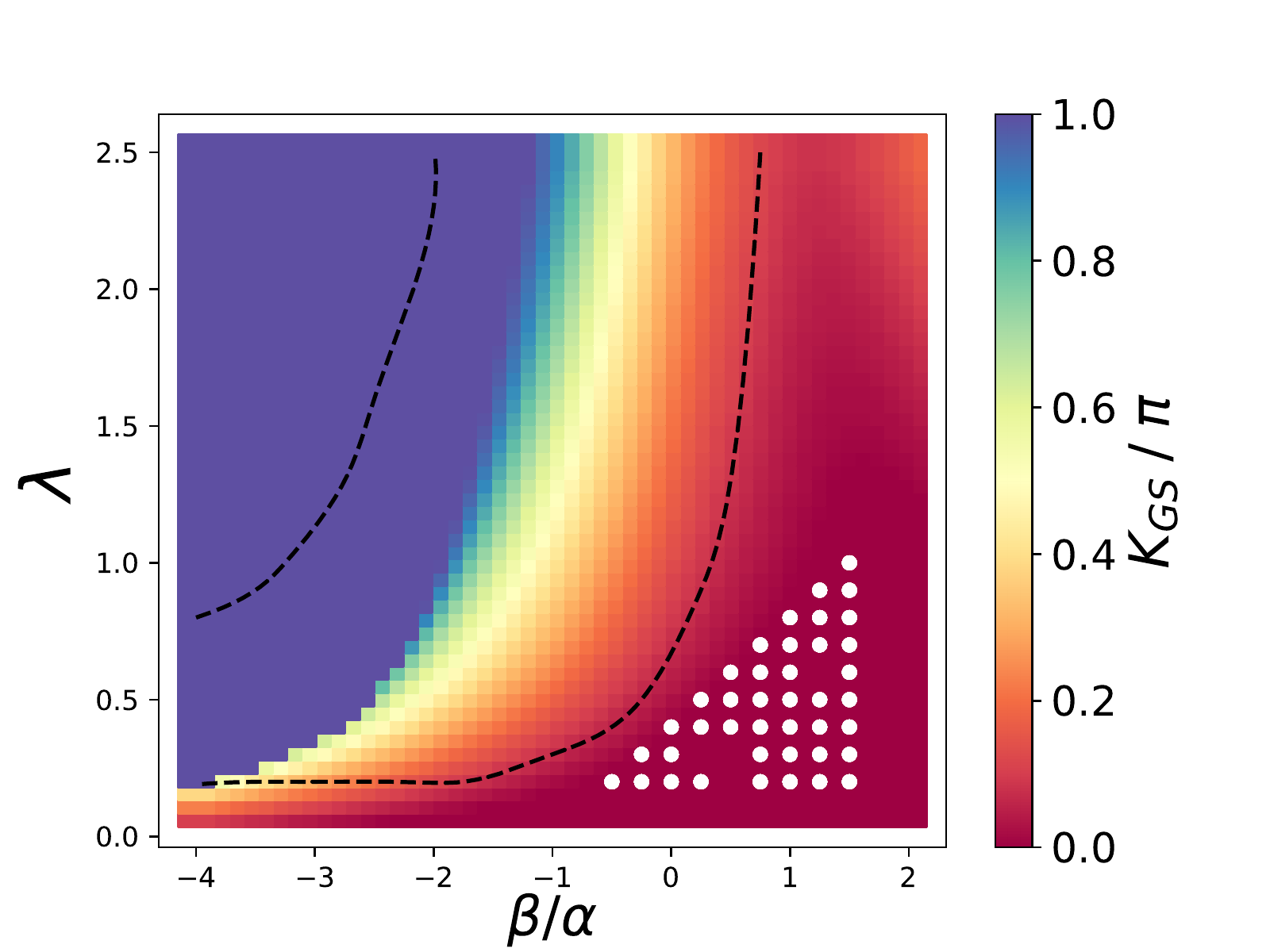}  
  \includegraphics[width=.45\textwidth]{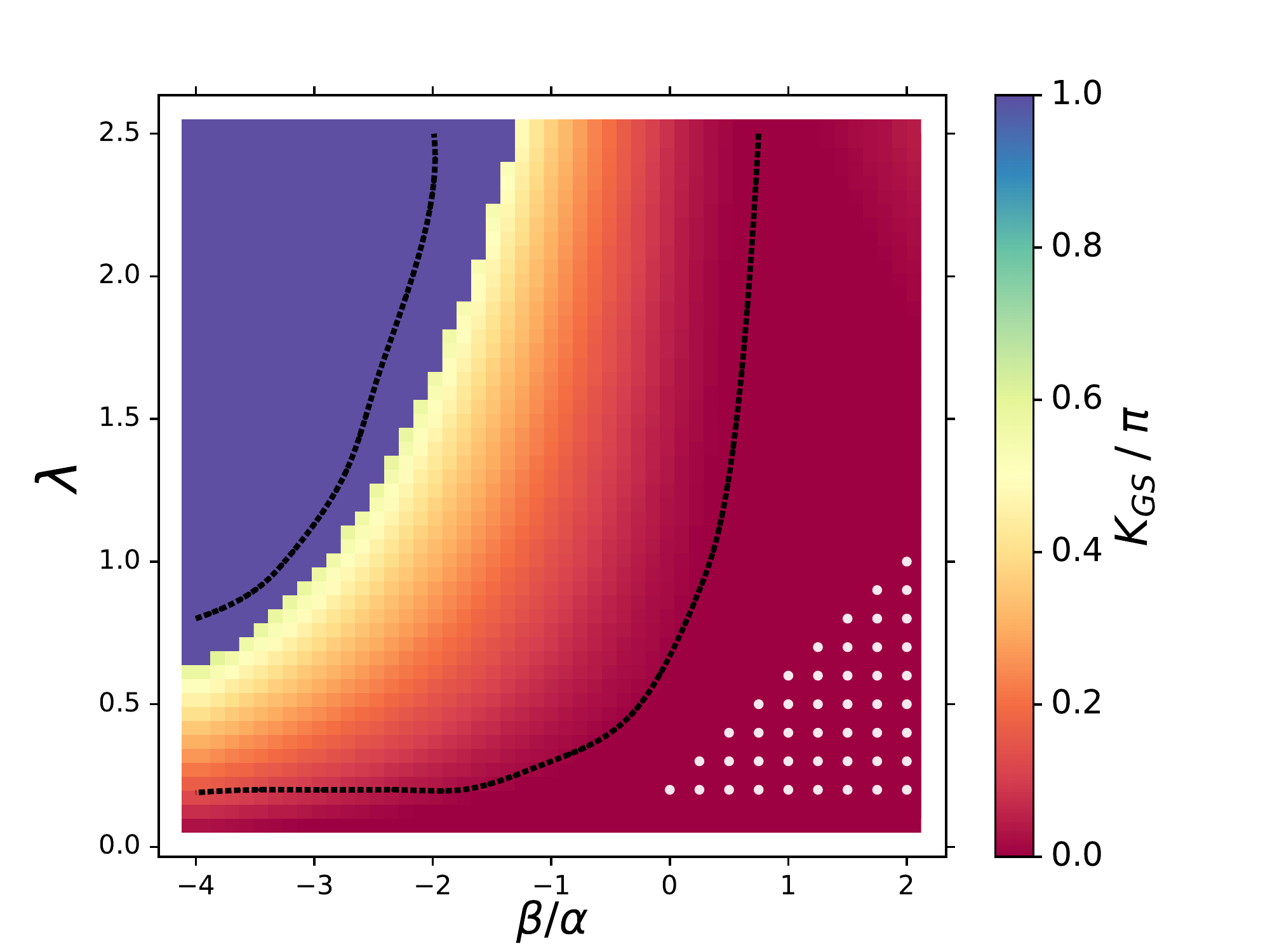}  
  \includegraphics[width=.45\textwidth]{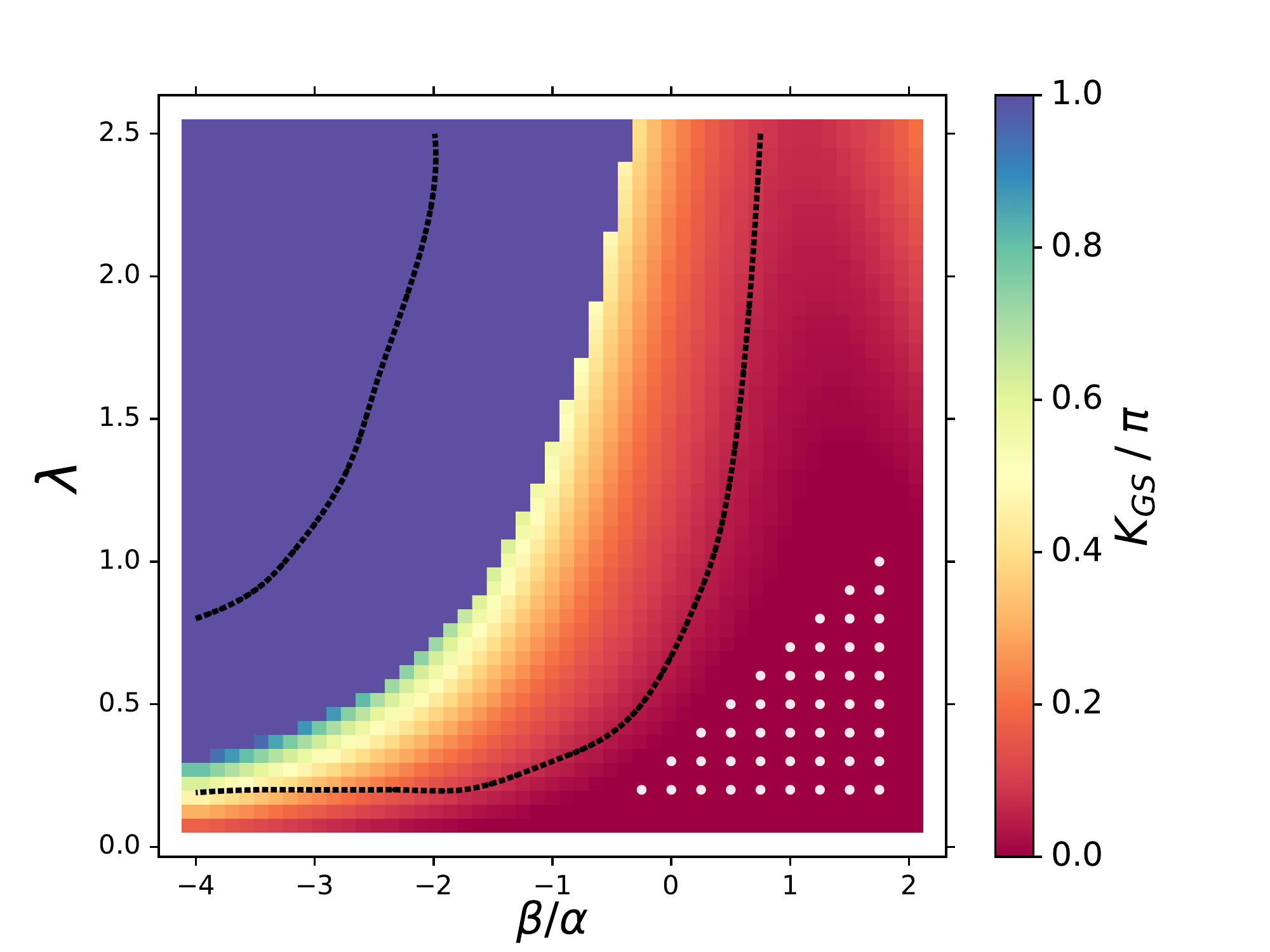}  
  \includegraphics[width=.45\textwidth]{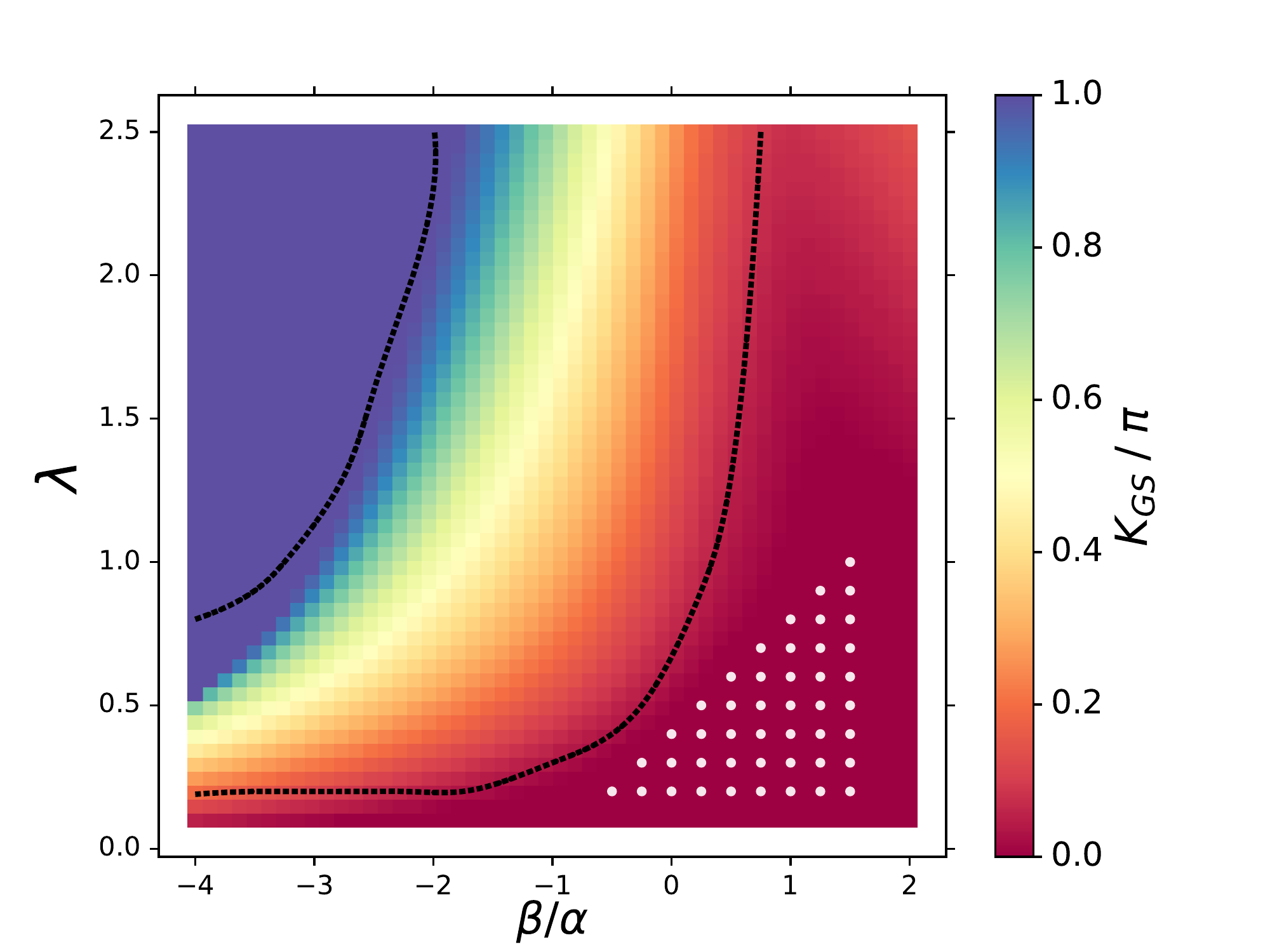}  
\caption{Dependence of the prediction accuracy on the number and positions of training points. 
The white dots indicate the values of the parameters $\lambda$ and $\alpha/\beta$, at which the quantum properties were calculated for training the GP models. All results are computed with optimal kernels with the same complexity level GPL-2.
}
\label{phases-training-points}
\end{figure}

Validation of the generalization predictions in the extrapolated region presents a major problem. By definition, there are no data in the extrapolated region. One can, of course, divide the given data into a training set and a validation set outside of the training data range. The validation set can then be used to verify the accuracy of the extrapolation. This is what is done throughout this work. However, this does not guarantee the accuracy of the predictions beyond the range of the validation data. 
Finding a proper method to validate the extrapolation predictions is particularly important for applications of the present approach to making predictions of observables at physical parameters, where no theoretical or experimental results are available. 

A possible way to verify the accuracy of the extrapolation predictions without using data in the extrapolated region is to examine the sensitivity of the predictions to the {\it positions} and  {\it number} of training points. If the predictions are stable to variations of the training data, one might argue that the predictions are valid.  To illustrate this, we  rebuild the models of the phase diagram depicted in Figure \ref{phases} with a variable number of training points. 
Figure \ref{phases-training-points} shows the results obtained with models trained by the quantum calculations at different values of $\lambda$ and $\alpha/\beta$. The figure illustrates the following: 

\begin{itemize}

\item[$\circ$] The generalization models capture both transitions even when trained by the quantum calculations far removed from the transition line and with a random distribution of training points. 

\item[$\circ$] The predictions of the transitions become more accurate as the distribution of the training points approaches the first transition line. 

\end{itemize}

\begin{figure}[h!]
\centering
  \includegraphics[width=.42\textwidth]{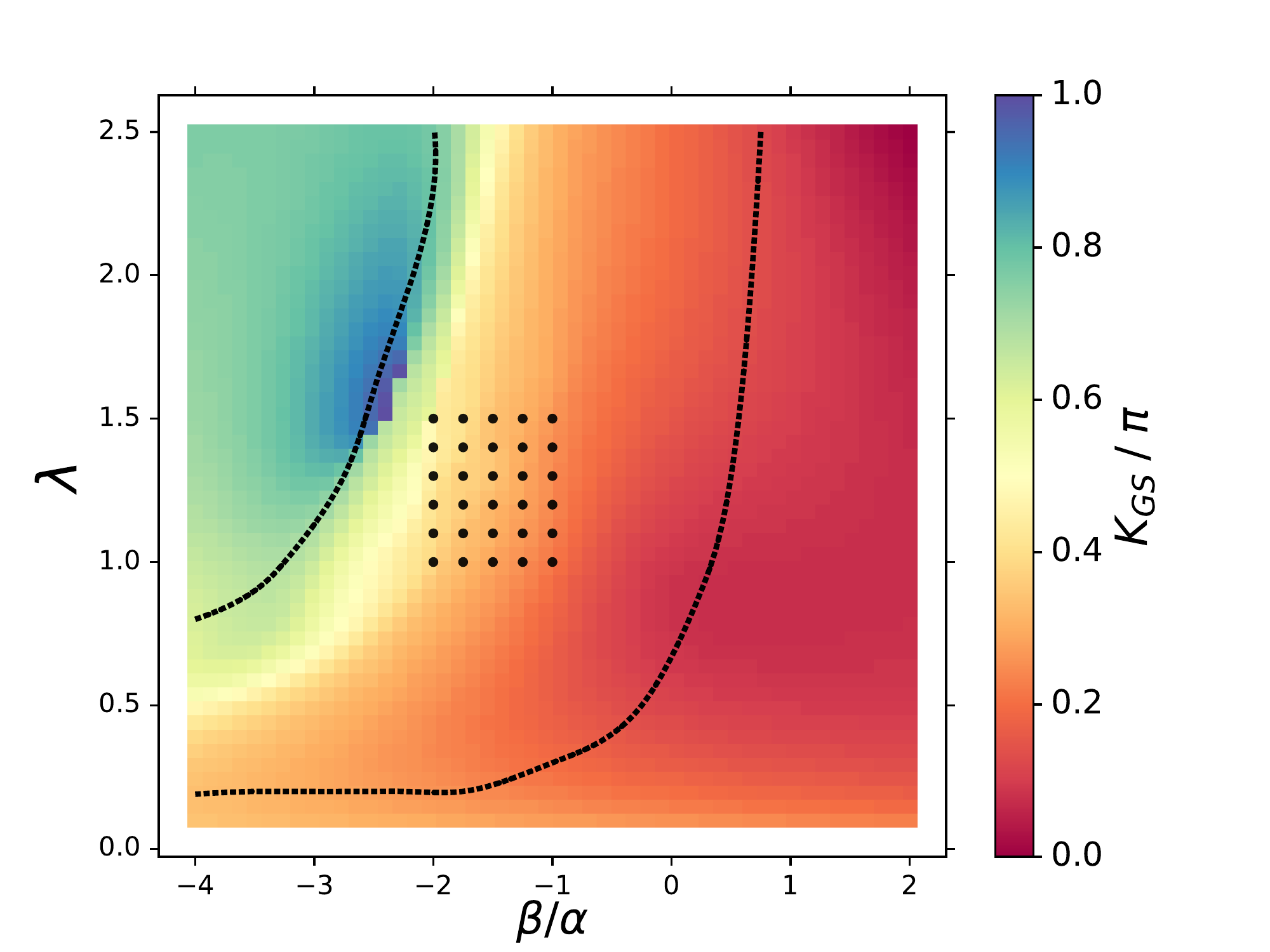}  
  \includegraphics[width=.42\textwidth]{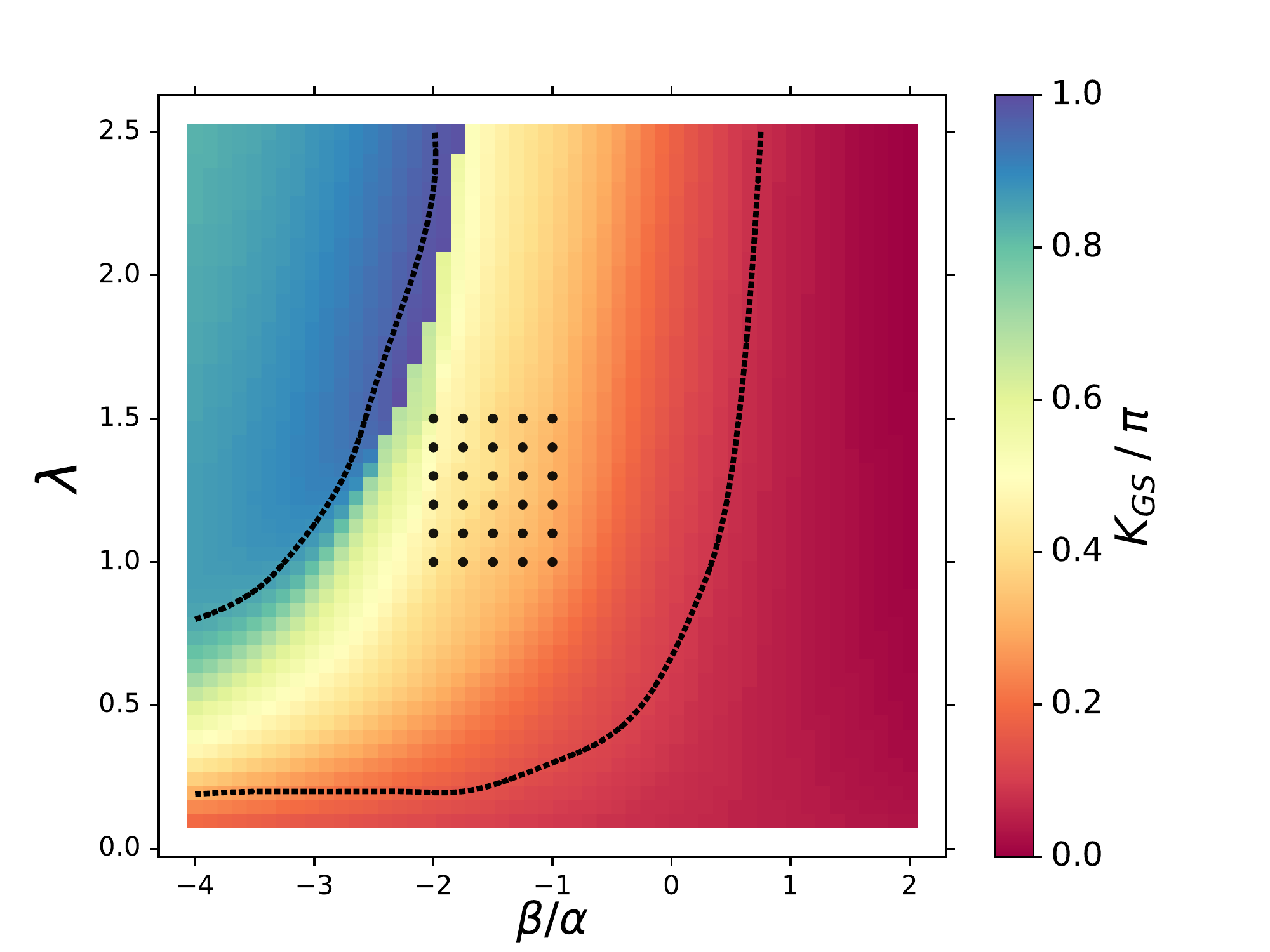}  
  \includegraphics[width=.42\textwidth]{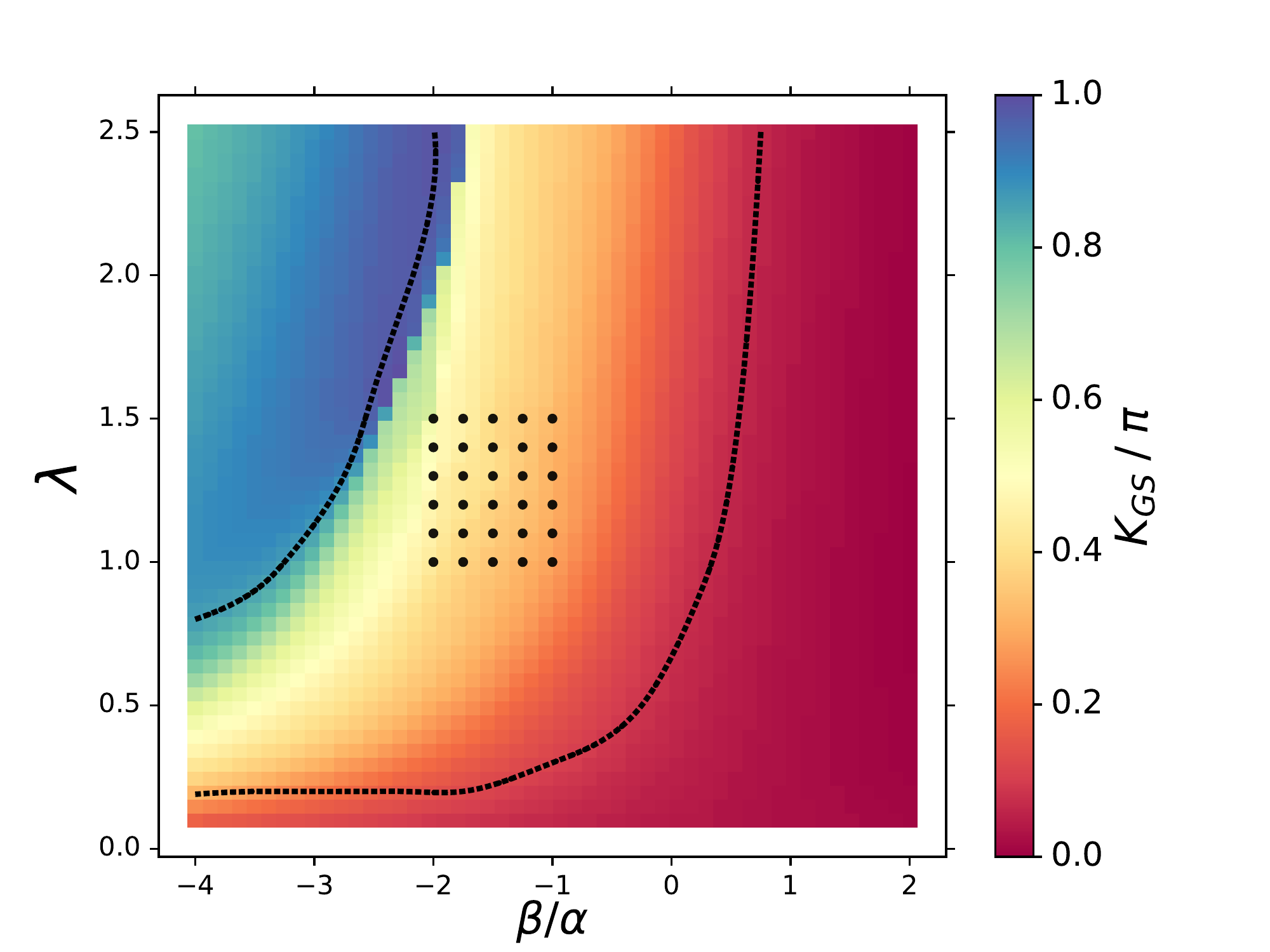}  
  \includegraphics[width=.42\textwidth]{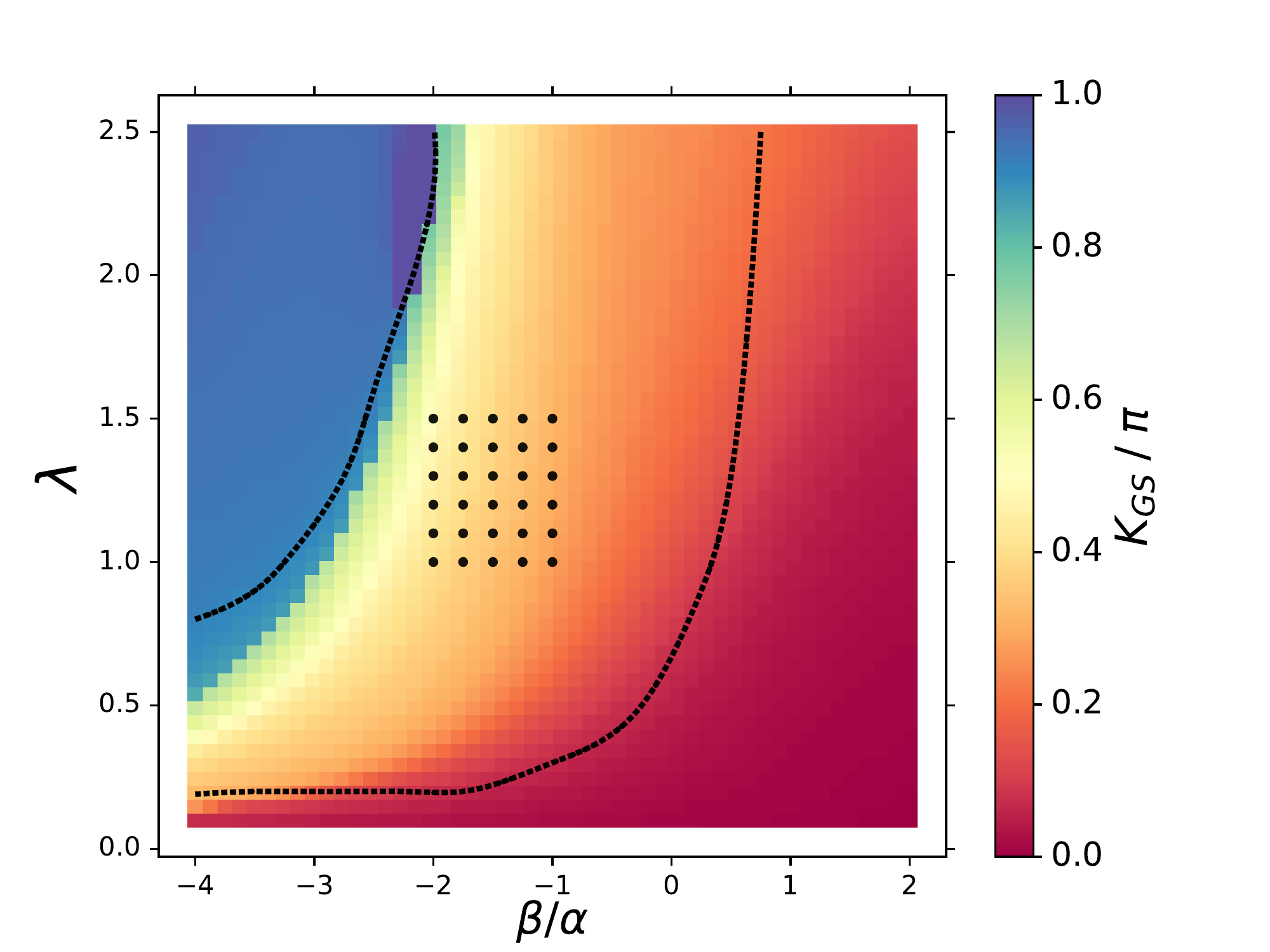}  
  \includegraphics[width=.42\textwidth]{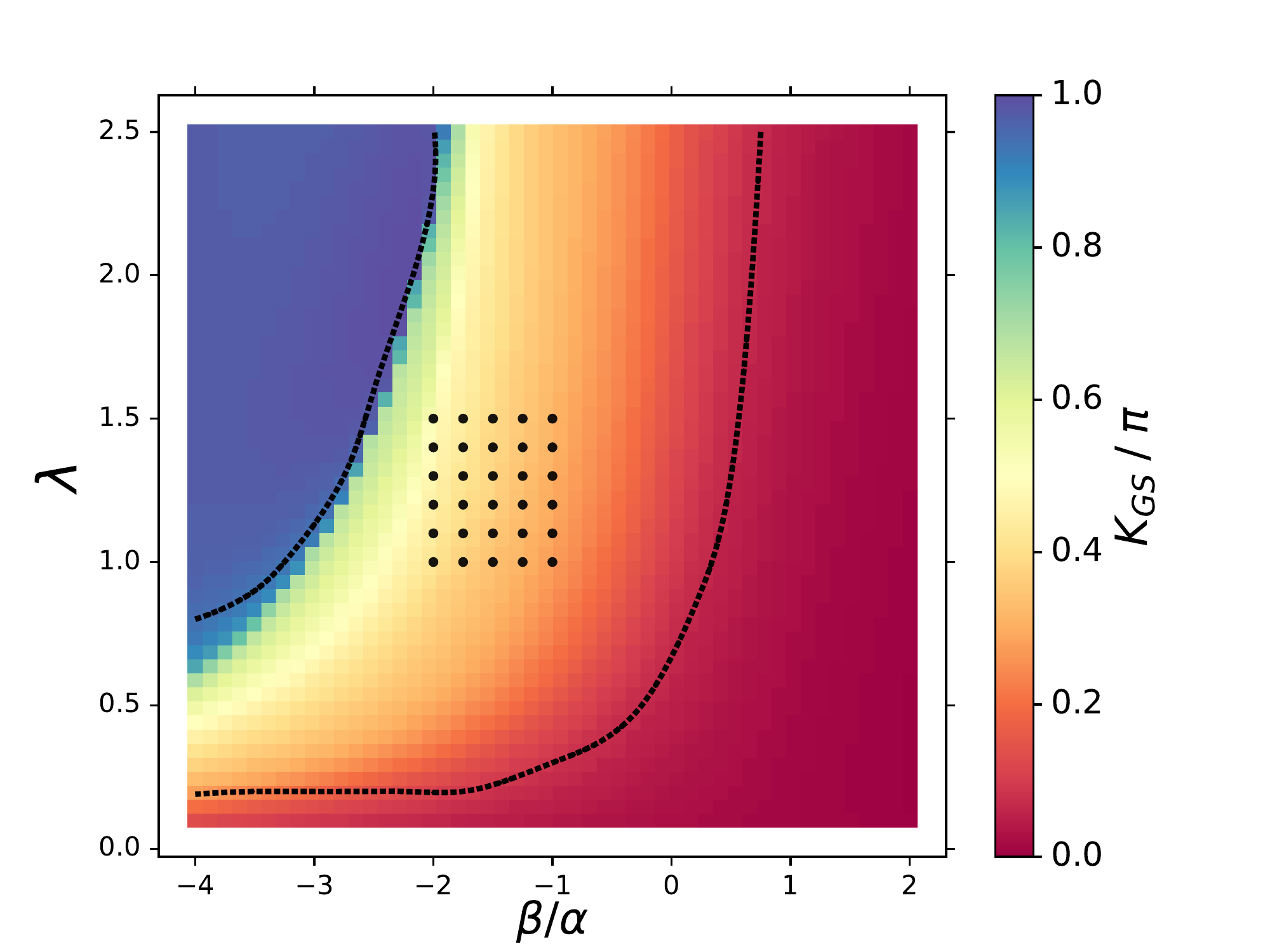}  
\caption{Dependence of the prediction accuracy on the positions of training points. 
The black dots indicate the values of the parameters $\lambda$ and $\alpha/\beta$, at which the quantum properties were calculated for training the GP models. The different panels correspond to the optimal kernels with the complexity level ranging from GPL-0 (upper left) to GPL-4 (lowest panel).
}
\label{phase2-training-points}
\end{figure}

One may thus conclude that the predictions of the sharp transitions are physical. If possible, this can be further validated by training generalization models with data in a completely different part of the phase diagram. This is done in Figure \ref{phase2-training-points} that shows the same phase diagram obtained by the generalization models trained with quantum results entirely in the middle phase. Given the remarkable agreement of the results in the last panel of Figure \ref{phases-training-points} and in the last panel of Figure \ref{phase2-training-points}, one can argue that the predicted phase diagram is accurate. \\

Based on these results, we suggest the following algorithm to make stable predictions of unknown phase transitions by extrapolation: 

 \begin{itemize}
\item[$(1)$] $\;$ Sample the phase diagram with a cluster of training points at random. 
\item[$(2)$] $\;$ Identify the phase transitions by extrapolation in all directions. 
\item[$(3)$] $\;$ Move the cluster of the training points towards any predicted transition. 
\item[$(4)$] $\;$ Repeat the calculations until the predictions do not change with the change of the training point distributions.  
\item[$(5)$] $\;$ If possible, rebuild the models with training points in a completely different part of the phase diagram.  
\end{itemize}
While step (5) is not necessary, the agreement of the results in steps (4) and (5) can be used as an independent verification of the extrapolation. The comparison of the results in steps (4) and (5) may also point to the part of then phase diagram, where the predictions are least reliable. 

\subsection{Power of the Bayesian Information Criterion}
\label{subsec:BIC2}

\begin{figure}
\centering
  \includegraphics[width=.42\textwidth]{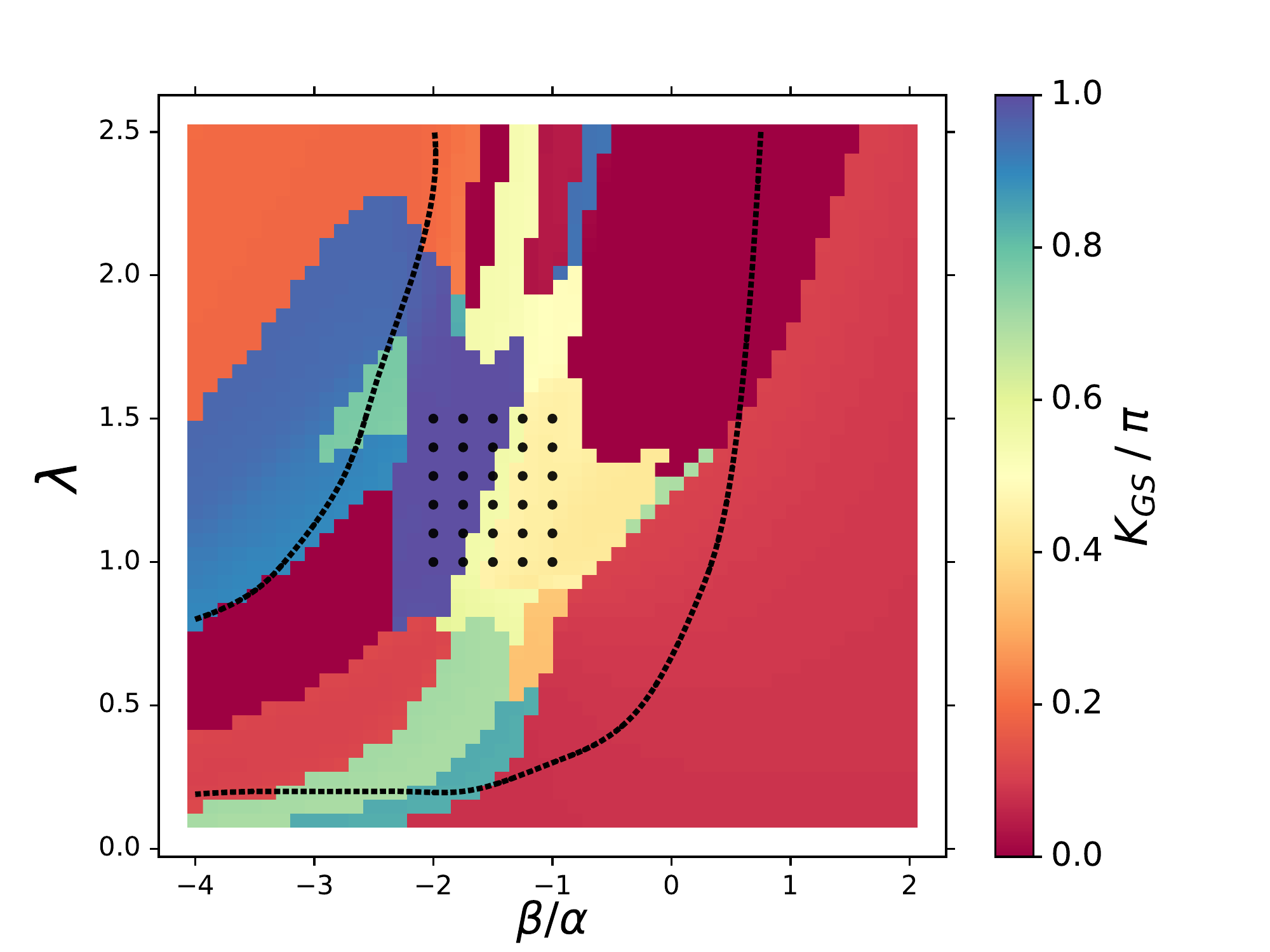}  
  \includegraphics[width=.42\textwidth]{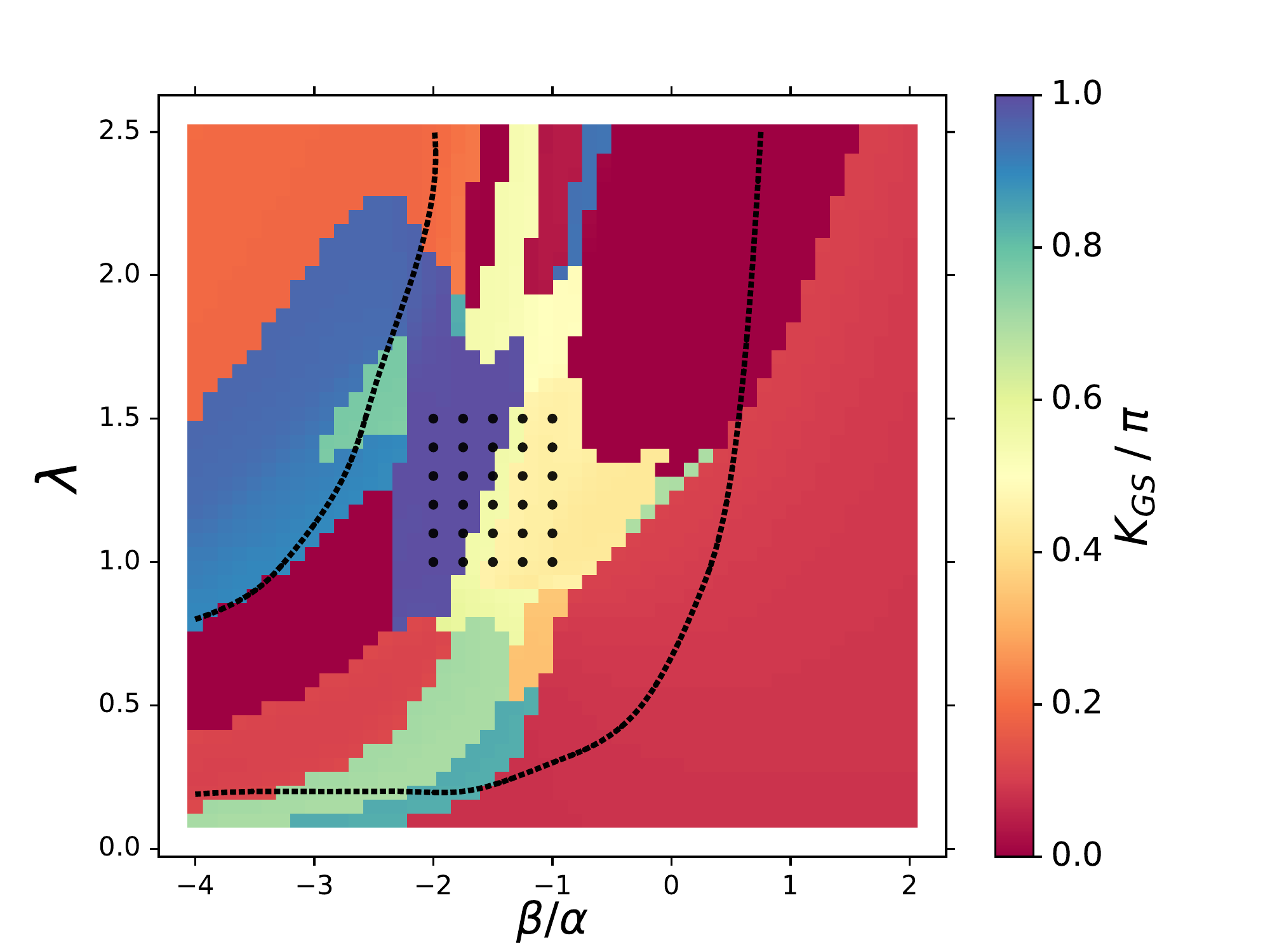}  
  \includegraphics[width=.42\textwidth]{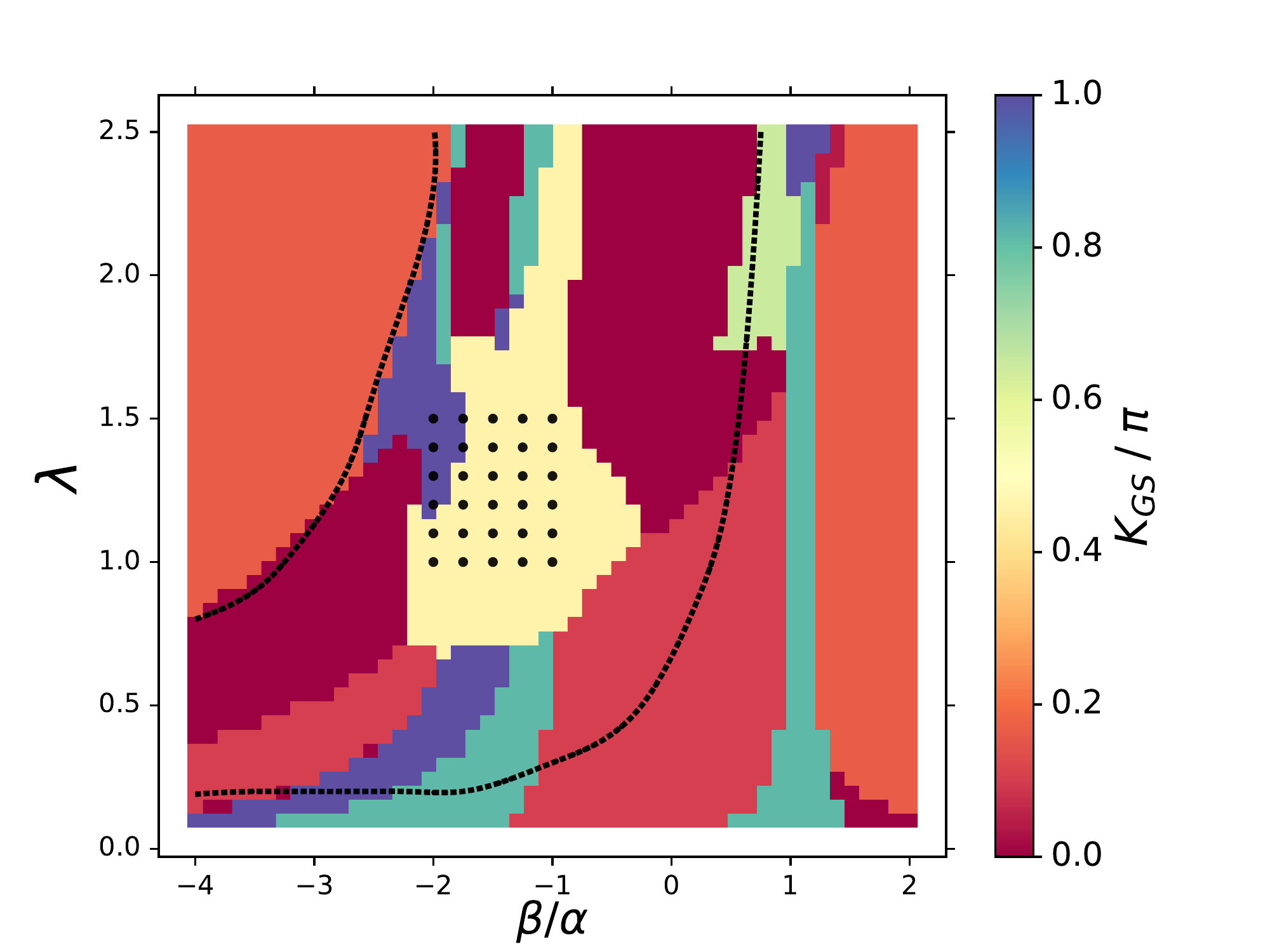}  
  \includegraphics[width=.42\textwidth]{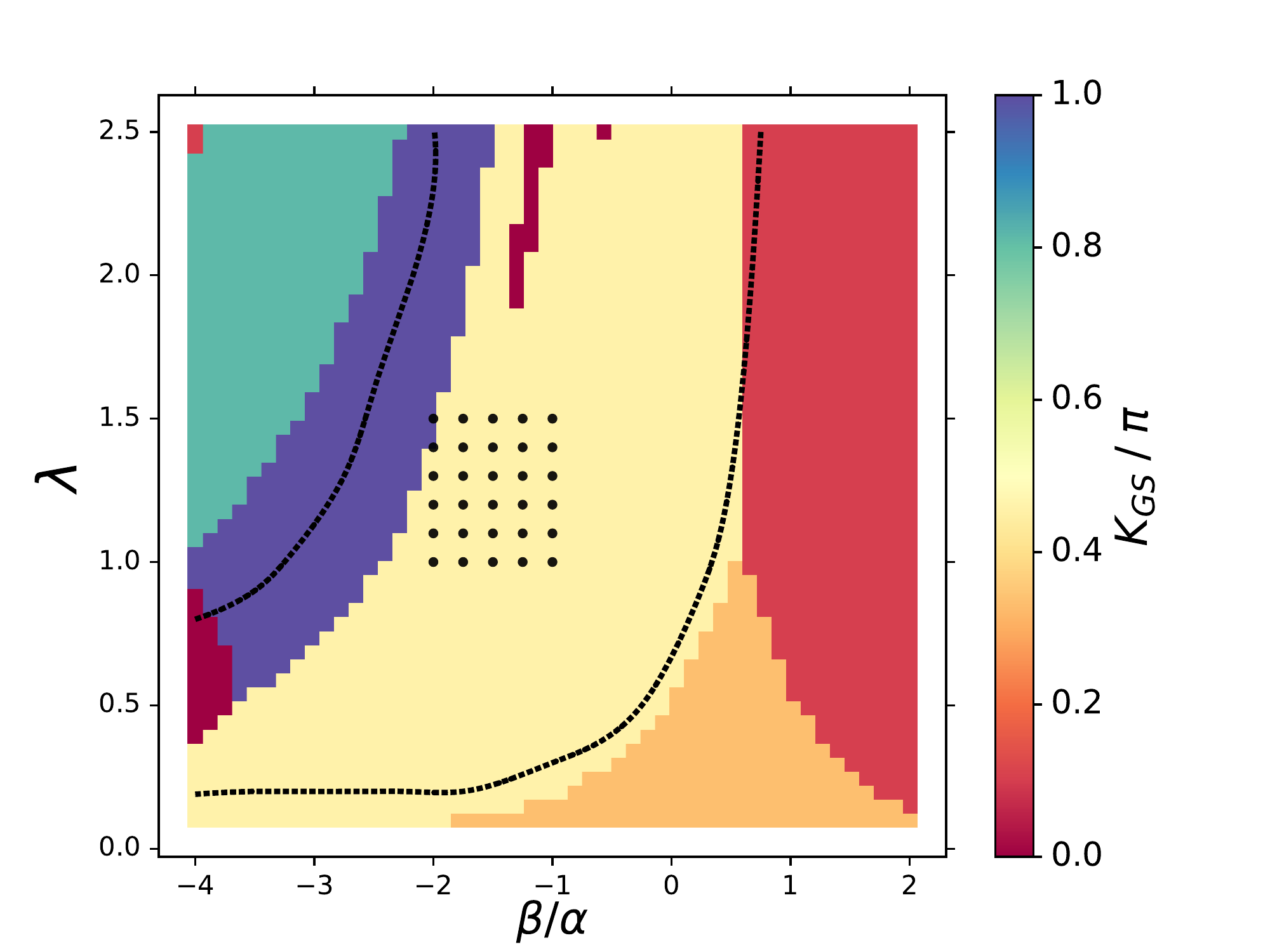}  
  \includegraphics[width=.42\textwidth]{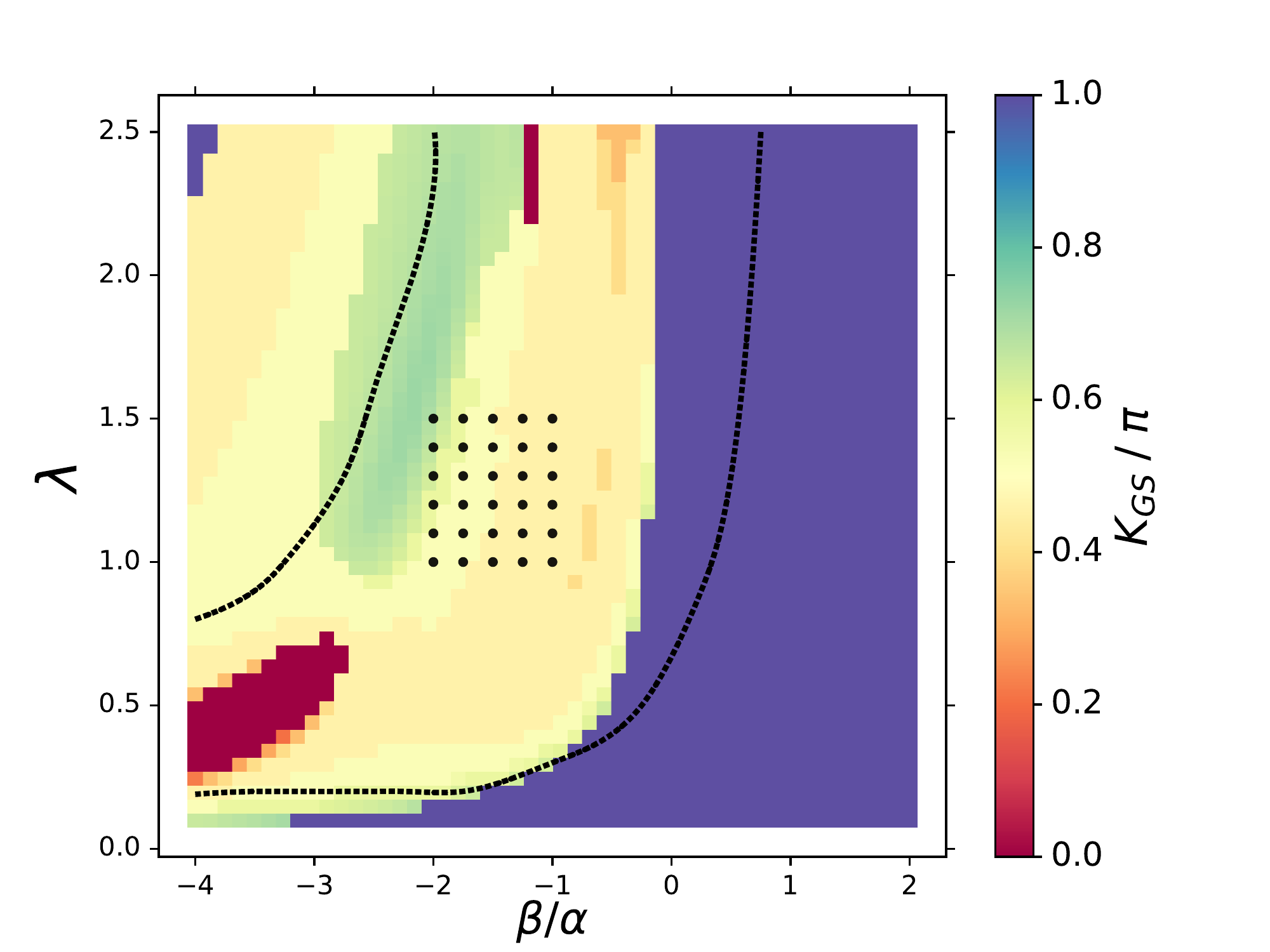}  
\caption{Predictions obtained with randomly selected kernels.  
The black dots indicate the values of the parameters $\lambda$ and $\alpha/\beta$, at which the quantum properties were calculated for training the GP models. The different panels correspond to the optimal kernels with the complexity level ranging from GPL-0 (upper left) to GPL-4 (lowest panel). The initial kernel is selected at random. The kernel at the next complexity level GPL-X is obtained by combining the kernel from the previous complexity level with another randomly selected kernel. The parameters of the kernels thus obtained are optimized using the maximization of the log marginal likelihood. This procedure illustrates the importance of the BIC for the selection of the type of the kernel function. 
}
\label{random-kernels}
\end{figure}

As explained in Section \ref{sec:gpss_algorithm}, the generalization models used here are obtaining by gradually increasing the complexity of kernels using the BIC (\ref{BIC-eq}) as a kernel selection criterion. The algorithm starts with a simple kernel that leads to a model with the largest BIC. This kernel is them combined with multiple simple kernels leading to multiple models. The kernel of the model with the largest BIC is selected as a new kernel, which is again combined with multiple simple kernels. The procedure is iterated to increase the kernel complexity, one simple kernel at a time. Since the BIC (\ref{BIC-eq}) is closely related to the log marginal likelihood and the kernel parameters are optimized for each step by maximizing the log marginal likelihood, why not to simply select some complex kernel function at random and maximize the log marginal likelihood of the model with this kernel? 

To illustrate the power of the BIC in the greedy search algorithm, we repeat the calculations of Figure \ref{phase2-training-points} with kernels of various complexity selected at random. We mimic the iterative process used above, but, instead of using the BIC as a kernel selection criterion, we select a new kernel at each complexity level at random. Every model is optimized by maximizing the log marginal likelihood as usual. The results are depicted in Figure \ref{random-kernels}. The different panels of Figure \ref{random-kernels} are obtained with models using kernels of different complexity. The models are not physical and there is no evidence of model improvement with increasing kernel complexity. We thus conclude that the BIC is essential as the kernel selection criterion to build GP models for applications targeting physical extrapolation.

\section{Conclusion}
\label{sec:conclusion}

The present article presents clear evidence that Gaussian process models can be designed to predict the physical properties of complex quantum systems outside the range of the training data. As argued in Ref. \cite{BML}, the generalization power of GP models in the extrapolated region is likely a consequence of the Bayesian approach to machine learning that underlies GP regression. 
For this reason, the authors believe that Bayesian machine learning has much potential for applications in physics and chemistry \cite{BML}. As illustrated here, it can be used as a new discovery tool of physical properties, potentially under conditions, where neither theory nor experiment are feasible.

The generalization models discussed here can also be used to guide rigorous theory in search of specific phenomena (such as
phase transitions) and/or particular properties of complex systems. Generating the phase diagram, such as the one depicted in Figure \ref{phases}, presents no computational difficulty (taking essentially minutes of CPU time). One can thus envision the following efficient approach for the generation of the full phase diagrams based on a combination of the GP models with rigorous calculations or experiments: 

\begin{itemize}

\item[$(1)$] $\;$ Start with a small number of rigorous calculations or experimental measurements. 

\item[$(2)$]  $\;$  Generate the full phase diagram with the GP models with complex kernels. This diagram is likely to be inaccurate at the system parameters far away from the initial training points. 

\item[$(3)$] $\;$ Use the rigorous calculations or experiments to add training points in the parts of the parameter space, where (a) the system exhibits desired properties of interest; and (b) where the system properties undergo the most rapid change.  

\item[$(4)$] $\;$ Repeat the calculations until the predictions in the extrapolated region do not change with the change of the training point distributions.  

\end{itemize}
With this approach, one can envision generating complete ${\cal D}$-dimensional phase diagrams with about $10 \times {\cal D}$ rigorous calculations or experimental measurements. Training the models and making the predictions in step (2) will generally take a negligibly small fraction of the total computation time. 

It should be pointed out that the results presented in this work suggest algorithms to construct complex GP models capable of meaningful predictions in the extrapolated region without direct validation. To do this, one can examine the sensitivity of the predictions to the distribution of the training points for models with the same level of kernel complexity as well as models with different complexity. Increase of the sensitivity to the training points with the kernel complexity would suggest overfitting or insufficient optimization of the kernel parameters. In such cases, the iterative process building up the kernel complexity should be stopped or the process of optimizing the kernel parameters revised. Constructing algorithms for physical extrapolation without the need for validation should be the ultimate goal of the effort aimed at designing ML models for physics and chemistry. Such models could then use all available chemistry and physics information to make meaningful discoveries. 

\section*{Acknowledgments}

We thank Mona Berciu for the quantum results used for training and verifying the ML models for the polaron problem. We thank John Sous and Mona Berciu for the ideas that have led to work published in Ref. \cite{extrapolation-paper} and for enlightening discussions.



\bibliography{reference}

\end{document}